\documentclass[journal,letterpaper]{IEEEtran}

\usepackage{amsmath,amssymb,amsfonts,amsthm,empheq}
\usepackage{graphicx} 
\usepackage{color}
\usepackage{hyperref}
\usepackage{dsfont}
\usepackage{bbm}
\usepackage[normalem]{ulem}

\newtheorem{theorem}{Theorem}
\newtheorem{lemma}{Lemma}

\newtheorem{definition}{Definition}

\newcommand{\beq}{\begin{equation}}
\newcommand{\eeq}{\end{equation}}
\newcommand{\bea}{\begin{eqnarray}}
\newcommand{\eea}{\end{eqnarray}}
\newcommand{\BO}{\mathcal{B}_{\Omega}}
\newcommand{\BQ}{\mathcal{B}_{\mathcal{Q'}}}  

\DeclareMathOperator*{\argmin}{arg\,min}

\begin{document}

\sloppy

\title{Deterministic coding theorems for   blind sensing:     optimal measurement rate and fractal dimension}


\author{
	Taehyung~J.~Lim and Massimo~Franceschetti,~\IEEEmembership{Senior Member, IEEE}  
	\thanks{The authors are with the Information Theory and Applications Center (ITA) of the California Institute of Telecommunications and Information Technologies (CALIT2), Department of Electrical and Computer Engineering, University of California, San Diego CA, 92093, USA. Email: taehyung.lim@hotmail.com, massimo@ece.ucsd.edu. This work was presented in part at the IEEE International Symposium on Information Theory, Aachen, Germany, June 2017. This work was partially supported by the National Science
Foundation under Award CCF-1423648.}  
}

\maketitle


\begin{abstract}
Completely blind sensing is the problem of  recovering bandlimited signals from measurements, without any spectral information beside an upper bound on the measure of the whole support set in the frequency domain.
Determining the  number of measurements necessary and sufficient for reconstruction has been an open problem, and usually partially blind sensing is performed,  assuming  to have some partial spectral information available a priori. In this paper,      the minimum number of measurements that guarantees perfect recovery in the absence of measurement error, and robust recovery in the presence of measurement error, is determined in a completely blind setting. Results  show that a factor of two in the measurement rate is the price pay for blindness, compared to reconstruction with  full spectral knowledge.  The minimum number of measurements is also related to the fractal (Minkowski-Bouligand) dimension  of a   discrete approximating set,  defined in terms of the Kolmogorov $\epsilon$-entropy.  These results are analogous to a   deterministic coding theorem, where an operational quantity defined in terms of minimum measurement rate  is shown to be equal to an information-theoretic one.  A comparison with parallel results in compressed sensing is   illustrated, where the relevant dimensionality notion in a stochastic setting is   the information (R\'{e}nyi) dimension, defined in terms of the Shannon entropy. 

\end{abstract}



\section{Introduction}
\subsection{Problem set-up}
\label{sec:intro}
Let $f: \mathbb{R} \rightarrow \mathbb{R}$ be square-integrable  and  such that
\beq
\mathfrak{F} f(\omega)=0, \mbox{ for } \omega \not \in \mathcal{Q},
\label{sigdef1}
\eeq
where $\mathfrak{F}$ indicates   Fourier transform, $\omega$ indicates angular frequency, and $\mathcal{Q} $ is a   subset of the interval $[-\Omega,\Omega]$  of measure
\beq
m(\mathcal{Q}) \leq  2 \Omega'. 
\label{sigdef2}
\eeq

\begin{figure}
\begin{center}
{
{\includegraphics[width=85mm]{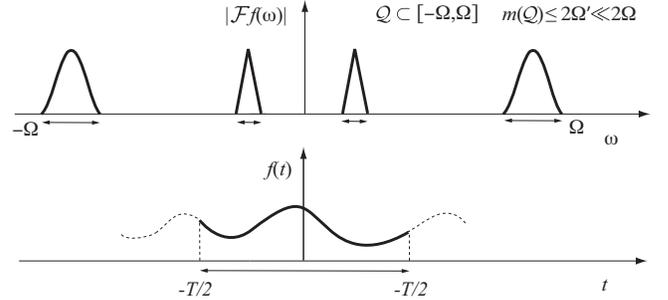}}}
\end{center}
\caption{Illustration of a sparse multi-band signal observed over a single time interval. }
\label{fig:multibandS}
\end{figure}

A typical example   occurs when $\mathcal{Q}$ is the union of a finite number of disjoint sub-intervals of $[-\Omega,\Omega]$ and $\Omega' \ll \Omega$,
see Figure~\ref{fig:multibandS}. 
These kind of signals arise  in many applications, ranging from radio,   to   audio,    and biological  communication and sensing systems. A natural question is what is the minimum number  of measurements  that can be performed   over a given time  interval and  that  guarantees reconstruction with a minimum amount of error.

To address this question,  
we consider a measurement vector ${\bf y} \in \mathbb{R}^M$
\beq
	{\bf{y}} = \mathfrak{M}f(t) + {\bf{e}}, 
	\label{mop}
\eeq
where $\mathfrak{M}$ is an  operator from multi-band signals  to $M$-dimensional vectors and ${\bf{e}} \in \mathbb{R}^M$ is the measurement error.

We assume each measurement $y_n\in \bf{y}$  results from observing the signal over the interval $[-T/2,T/2]$ through the inner product with a  bandlimited kernel, plus an error term. 
\begin{definition} \emph{(Measurements)}
For all   $n \in \{1,\ldots, M\}$, we have
\begin{align}
	y_n & = \int_{-T/2}^{T/2} f(t) \varphi_n(t) dt  + e_n, \;\;\;   
	\label{singlemeasurement}
\end{align} 
where
\begin{align}
\mathcal{F}\varphi_n(\omega) = 0 \mbox{ for } \omega \not \in [-\Omega,\Omega].
\end{align}
\end{definition}

\begin{figure}[t] 
\begin{center}
{
{\includegraphics[width=80mm]{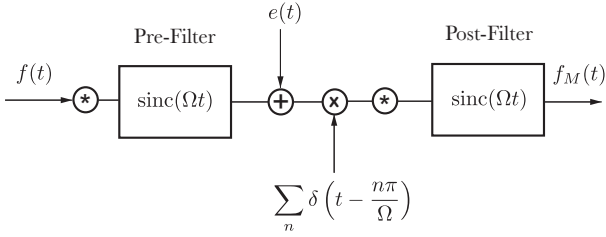}}}
\end{center}
\caption{Block diagram for sampling measurement and reconstruction. The symbol $*$ indicates convolution. }
\label{fig:sampling}
\end{figure}

\begin{figure}
\begin{center}
{
{\includegraphics[width=80mm]{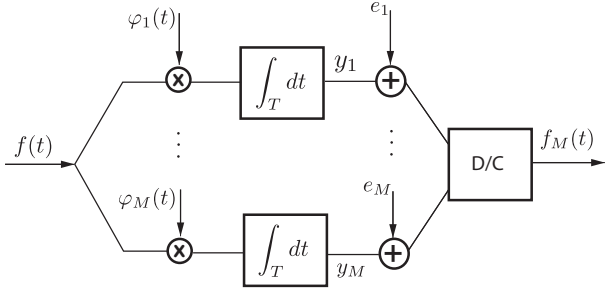}}}
\end{center}
\caption{Block diagram  for general measurement and reconstruction. The box D/C stands for discrete-to-continuous transformation and performs the  reconstruction of the signal from the discrete measurements. 
}
\label{fig:optimal1}
\end{figure}

This set-up  covers a wide range of real measurements.
Possible bandlimited kernels that fall in this framework include the Shannon cardinal basis $\mbox{sinc} (\cdot)$ functions~\cite{Jerri},
the Slepian prolate spheroidal wave functions (PSWF)~\cite{SlepianS1}, 
as well as other bandlimited functions  of practical interest, such as wavelets, and splines. The measurements are functionals of the signal over the entire observation interval, but in some cases they can reduce to the sampled signal values. For example,   for  
the cardinal basis  the measurements in (\ref{singlemeasurement}) also correspond to  low-pass filtering and sampling, and the signal can be   recovered by  low-pass filtering    the sampled signal values~\cite{unser}. This special case is illustrated in Figure~\ref{fig:sampling}.  The general case is illustrated in Figure~\ref{fig:optimal1}.

In the general setting, our aim is to determine  the smallest measurement rate 
\beq
	\bar{M} = \lim_{T \rightarrow \infty} \frac{M}{T}
	\label{rate}
\eeq
for which it is possible to obtain an approximation $f_M$  of $f$ from $\bf{y}$, such that the energy of the reconstruction error     is at most proportional to the energy of the measurement error, as  the size of the observation interval $T \rightarrow \infty$. This  corresponds to determining the scaling of the minimum number of measurements $M=M(T)$ that guarantees \emph{robust recovery} of any multi-band signal, namely a small perturbation in the measurement does not lead to a large reconstruction error.

\begin{definition}
\emph{(Robust recovery)}. There exists a universal constant $c \geq 0$, such that for $T$ large enough 
\begin{align}
\| f - f_M \|^2 & = \int_{-T/2}^{T/2} [f(t)-f_M(t)]^2 dt  \nonumber \\
& \leq c \sum_{n=1}^{M} e^2_n.
\end{align}
\end{definition}

When the measurement error tends zero, robust recovery reduces to   \emph{perfect recovery} of the signal. Namely, 
\begin{definition}
\emph{(Perfect recovery)}. 
\begin{align}
\lim_{T \rightarrow \infty} \| f - f_M \|^2 & = 0.
\end{align} 
\end{definition}

\subsection{Bandlimited signals}
Since   our signals are assumed to be bandlimited to $\Omega$,  one may readily observe that  in the absence of measurement error they can be perfectly recovered from a number of measurements slightly above the  Nyquist number  
\beq
N_0=\Omega T/\pi.
\eeq

For any $f$ satisfying (\ref{sigdef1}) and (\ref{sigdef2}),  and $\nu>0$,  we can construct an approximation $f_N$ of $f$  from a measurement vector ${\bf y}$ of size
\beq
N=(1+\nu) \Omega T/\pi, 
\label{eq:enne}
\eeq
and  such that  
\beq
\lim_{T \rightarrow \infty}  \| f - f_N\|^2 =0.
\label{eq:erra}
\eeq
This classic result is equivalent to stating that a measurement rate strictly above $\Omega/\pi$   is sufficient for reconstruction of any bandlimited singnal, and constitutes one of the milestones of electrical and communication engineering.

\begin{figure}
\begin{center}
{
{\includegraphics[width=70mm]{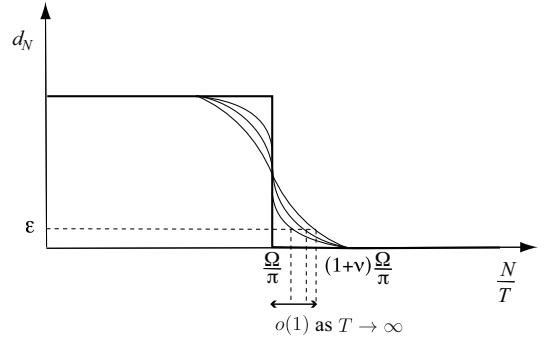}}}
\end{center}
\caption{Phase transition of the Kolmogorov $N$-width $d_N$ of bandlimited signals.  }
\label{kwidth}
\end{figure}

For  bandlimited signals,  the rate $\Omega/\pi$ is also optimal, in the following approximation-theoretic sense.
Consider performing
signal reconstruction by a linear interpolation of a number $N>0$ of orthogonal basis functions
\beq
f_N(t) = \sum_{n=1}^N y_n \varphi_n(t), 
\label{eq:prolaterep1}
\eeq
 and 
 let the Kolmogorov $N$-width    be  the smallest approximation error achievable  for all signals in the space, over all possible choices of basis sets. This minimum  error is achieved by measurements  that provide  the coefficients  of  the interpolation through the integrals
\beq
y_n = \frac{1}{\lambda_n} \int_{-T/2}^{T/2} f(t) \varphi_n(t) dt, \;\;\;  n \in \{1, \ldots, N\},
\label{eq:prolaterep2}
\eeq
where $\{\lambda_n\}$  are the eigenvalues of a Fredholm  integral equation of the second kind arising from Slepian's concentration problem~\cite{SlepianS1}, and the basis functions  $\{\varphi_n\}$ are  the corresponding eigenfunctions,   called PWSF~\cite{Flammer}.
The  measurement rate
   $\Omega/\pi$    corresponds to the critical threshold at which  the Kolmogorov $N$-width   transitions  from strictly positive values to zero, as $T \rightarrow \infty$~\cite{pinkus}.     This phase transition behavior of the approximation error is illustrated in Figure~\ref{kwidth}. 
With a number of measurements   $(1+\nu) \Omega T$ the error tends to zero as $T \rightarrow \infty$, while
with a number of measurements   $\Omega  T/\pi + o(T)$ the error remains positive as   $T \rightarrow \infty$.




\subsection{Multi-band signals}
For bandlilmtied signals that are supported over disjoint sub-bands,  an important extension of the  results above,  due to Landau and Widom~\cite{Landau1980}, states that  if we have a priori knowledge of the size and positions of all the sub-bands,    then signal reconstruction with vanishing error as $T \rightarrow \infty$ is   also possible using the smaller   number of measurements
\beq
S=(1+\nu)\Omega'T/\pi.
\eeq

A simple way to achieve this result  is  to demodulate each sub-band down to baseband, isolate it through low-pass filtering, and then sample each sub-band separately. The key contribution of Landau and Widom  is to  consider  the optimal subspace approximation, and show a  phase transition of the    error expressed in terms of Kolmogorov $N$-width. As in the single-band case,  a subspace approximation with vanishing error for all multi-band signals of a given frequency allocation  is obtained with a number of measurements $(1+\nu) \Omega'T$, while
a subspace approximation with vanishing error    is not possible for all multi-band signals using a number of measurements   $\Omega' T/\pi + o(T)$, and the value of the error is controlled by   the pre-constant in the $o(T)$ term.
It follows  that  for multi-band signals the Nyquist number $N_0=\Omega T/\pi$ can be replaced by the ``sparsity number" 
\beq
S_0=\Omega'T/\pi,
\label{sn}
\eeq
and the  \emph{occupied portion} of the frequency bandwidth   determines the   critical measurement rate $\Omega'/\pi$ required for reconstruction. 
In the case of sampling measurements,  Landau~\cite{Landau1967} also showed that a rate $\Omega'/\pi$ is  necessary for reconstruction, regardless of the reconstruction strategy being linear or not.

The results of above rely on two critical assumptions. First, they need a priori knowledge of the spectral occupation, since the eigenvalues and the optimal eigenfunctions used for reconstruction  are    solutions of an integral equation that depends on the spectral support set. In practice,  it might be difficult to know the exact number of sub-bands, their location,  and their widths prior to the measurements. A second critical assumption  is the absence of measurement error. In practice, the measurement process always carries a certain amount of error and its impact on the reconstruction error should be taken into account.

\subsection{Completely blind sensing}
In this paper, we  consider robust signal reconstruction   in the presence of    measurement error and without any a priori knowledge of the sub-bands beside an upper bound on the measure of the whole support set of the signal in the frequency domain. We call this   \emph{robust, completely blind sensing.}  
The blindness requirement is   important when detecting the sub-bands is impossible or too expensive to implement. The robustness requirement is important to guarantee stability in the reconstruction process.

Partially blind sensing,  where some partial spectral information is assumed, has been studied extensively. 
First key results were given in a series of papers by Bresler and co-authors~\cite{Feng1996, Bresler1996, Venkataramani1998}.  Later extensions \cite{Mishali2009, Mishali2010} reduced the number of a priori assumptions, but  still require  knowledge of the number of sub-bands, and of their widths. 
The same assumptions are made in
\cite{Izu2009, Sejdic2008, Davenport2012}.
The main result   in this setting is that the price to pay for partial blindness is   a factor of two in the  measurement rate. Several reconstruction strategies have been proposed using a   measurement rate above $2\Omega'/\pi$,
all  assuming some partial spectral knowledge, and lacking an information-theoretic converse. 
We remove these assumptions,   show that  a measurement rate $2 \Omega'/\pi$ is  sufficient for robust reconstruction in a completely blind setting,  and  provide a tight converse result. We also provide a deterministic  coding theorem for continuous analog sources, giving an interpretation of the minimum number of measurements in terms of the  ``effective" Minkowski-Boulingand dimension   of the infinite-dimensional set of multi-band signals, expressed in terms of the Kolmogorov $\epsilon$-entropy. This is compared with an analogous interpretation arising in the framework of compressed sensing, where the objective is the lossless source coding of a discrete, analog, stochastic process~\cite{Wu2010, Wu2012}. In that case, an analogous coding theorem has been given   in terms of  the R\'{e}ny dimension, expressed in terms of the Shannon entropy.

Finally, we remark  that while in the case of   multi-band signals of a given sub-band allocation the results of Landau and Widom     provide  an optimal subspace approximation in terms of  a linear interpolation of eigenfunctions  supported over multiple sub-bands, and having the highest energy concentration over the observation domain,  our results only provide an answer to the question of whether recovery is possible or not, without giving an explicit approximation procedure. In our case, the   discrete-to-continuous block  in Figure~\ref{fig:optimal1}  remains unknown.
Nevertheless, from an information-theoretic perspective  one is  primarily interested in the possibility of recovery using any discrete to continuous transformation, and does not wish to restrict reconstruction to a linear approximation strategy.  The explicit construction of practical blind recovery strategies is certainly of interest, and these should be compared with  the information-theoretic optimum determined here.

The rest of the paper is organized as follows: 
In section II we  describe our results. In section III we compare our results with  compressed sensing and illustrate coding theorems in deterministic and stochastic settings.  In section~IV we provide some definitions and   preliminaries   that are useful for our derivation. Proofs are given in section V and VI. Section VII draws conclusions and discusses future work.

\section{Description of the results}
\label{description}

\subsection{Noiseless Case}

\begin{theorem}
\label{thm1}
\emph{(Direct)}. 
In the absence of measurement error,  we can perfectly recover any signal $f$ satisfying (\ref{sigdef1}) and (\ref{sigdef2})	using a  measurement rate 
	\beq
		\bar{M}  >   \frac{2\Omega'}{\pi}.
	\eeq
	
\end{theorem}
\begin{theorem}
\label{thm2}
\emph{(Converse)}. In the absence of measurement error,  we cannot perfectly recover all signals $f$ satisfying (\ref{sigdef1}) and (\ref{sigdef2})
using a  measurement rate 
	\beq
		\bar{M} \leq   \frac{2\Omega' }{ \pi}.
	\eeq
\end{theorem}

These results can    interpreted in terms of  the effective dimensionality  of the signals' space, leading to a coding theorem. 
For bandlimited signals, the effective number of dimensions can be  identified with the Nyquist number $N_0=\Omega T/\pi$. For multi-band signals for which the location  and widths of all the sub-bands  is fixed a priori,  as in the Landau-Widom case,  it  can be identified with the sparsity number $S_0=\Omega'T/\pi$.  On the other hand, without any a priori  knowledge,  we need  to account for the additional degrees of  freedom of allocating the sub-bands in the frequency domain, and  our results indicate that the effective dimensionality increases to $2S_0$.

To make these considerations precise, we consider an information-theoretic quantity that measures the   dimensionality of a set in metric space, namely its fractal (Minkowski-Bouligand) dimension, which corresponds to the rate of growth of the Kolmogorov $\epsilon$-entropy of successively finer discretizations of the space, and represents the degree of fractality of the set~\cite{Falconer}.  

\begin{definition} \label{fractaldimension}
\emph{(Fractal dimension)}. 
For any subset $\mathcal{X}$ of a metric space, the fractal dimension  is  
\beq
\mbox{ \normalfont{dim}}_F(\mathcal{X}) = \lim_{\epsilon \rightarrow 0} \frac{H_\epsilon(\mathcal{X})}{-\log \epsilon},
\label{eq:defrenid}
\eeq
where $H_\epsilon$ is the Kolmogorov $\epsilon$-entropy~\cite{kolmo}.  
\end{definition}
If this limit does not exist, then the corresponding upper and lower fractal dimensions are defined using lim sup and lim inf, respectively.

We also define the dilation
\begin{definition}
\emph{(Minkowski  sum).}
\beq
\mathcal{X} \oplus \mathcal{X} = \{ {\bf x_1} + {\bf x_2} : {\bf x_1}, {\bf x_2} \in \mathcal{X} \}.
\label{eq:minsum}
\eeq
\end{definition}

%
Consider now  the set of all  bandlimited signals whose energy is at most one. These signals can be approximated by 
an infinite set  $\mathcal{X}_{\text{B}}$  of vectors, each containing  $N=(1+\nu) \Omega T/\pi$   real coefficients. Using the PSWF  as a basis for interpolation, every assignment of coefficients satisfying the given energy constraint approximates, with vanishing error as $T \rightarrow \infty$,   a bandlimited signal. 
In the appendix, we show that 
\beq
\label{appendix1}
	\mbox{dim}_F(\mathcal{X}_{\text{B}}) = \mbox{dim}_F(\mathcal{X}_{\text{B}} \oplus \mathcal{X}_{\text{B}}),
\eeq
and  letting the fractal dimension rate of the approximating set be
\beq
R_F(\mathcal{X}_{\text{B}}) =	\lim_{T \rightarrow \infty}  \frac{\mbox{dim}_F(\mathcal{X}_\text{B}) }{T},
\eeq
we have
\beq
\label{appendix2}
R_F(\mathcal{X}_{\text{B}}) =\Omega/ \pi,
\eeq
which coincides with the measurement rate needed for reconstruction.
%

Next, we quantize the bandwidth at level $\Delta>0$  and let 
\beq
\mathcal{J} = \{ -\Omega, -\Omega + \Delta, -\Omega + 2\Delta, \cdots, \Omega \}.
\label{mathD}
\eeq
We consider the
subset of all  multi-band signals of a given   sub-band allocation,   whose energy is at most one, and such that     the extremal points of all sub-bands belong to   $\mathcal{J}$.  This subset of signals approximates, with vanishing energy error  as   $\Delta \rightarrow 0$, the one of all multi-band signals of a  given sub-band allocation and of energy at most one. It can also    be approximated, with vanishing error as $T \rightarrow \infty$,  by  an infinite set  $\mathcal{X}_{\text{MB}}(\Delta)$  of vectors, each containing  $N=(1+\nu) \Omega T/\pi$   real coefficients of a PSWF interpolation. Compared to the previous case, the choice of   the coefficients is  now restricted by the given sub-band allocation, so that   we have 
\beq
\mathcal{X}_{\text{MB}}(\Delta)  \subset \mathcal{X}_\text{B}. 
\eeq
Following the same argument used to derive \eqref{appendix1}, we   obtain
\beq
\label{appendix3}
	\mbox{dim}_F [\mathcal{X}_{\text{MB}}(\Delta)]  = \mbox{dim}_F [\mathcal{X}_{\text{MB}}(\Delta) \oplus \mathcal{X}_{\text{MB}}(\Delta)].
\eeq
In this case, however,   the   $N$-dimensional prolate spheroidal approximation is somewhat redundant, and following the same argument used to derive \eqref{appendix2}, we obtain
\beq
\label{appendix4}
	\lim_{\Delta \rightarrow 0}  R_F[\mathcal{X}_{\text{MB}}(\Delta)] =\Omega'/ \pi,
\eeq
which coincides with the Landau-Widom rate~\cite{Landau1980, Landau1967} needed for reconstruction.
%

Finally, consider the subset of all multi-band signals whose energy is at most one, having an arbitrary sub-band allocation of measure at most $2\Omega'$, and  such that    the extremal points of all sub-bands belong to $\mathcal{J}$. These   signals   can be approximated, as $T \rightarrow \infty$,  by 
an infinite set  $\mathcal{X}(\Delta)$  of vectors, each containing  $N=(1+\nu) \Omega T/\pi$   real coefficients of a PSWF interpolation. The choice of the coefficients is  now restricted only by the measure of the occupied portion of the spectrum and not by a specific sub-band allocation,    and we have 
\beq
\mathcal{X}_{\text{MB}}(\Delta)  \subset \mathcal{X}(\Delta) \subset \mathcal{X}_\text{B}.
\eeq
By combining Theorems~\ref{thm1} and \ref{thm2} with Theorems \ref{thm3} and \ref{thm4} below,  we obtain
\beq
	\label{appendix6}
	\lim_{\Delta \rightarrow 0}  R_F[\mathcal{X}(\Delta)]  =  \Omega'/ \pi.
\eeq


%
%
\begin{theorem}
\label{thm3} 
\emph{(Direct)}. In the absence of measurement error,   we can perfectly recover any signal $f$ satisfying (\ref{sigdef1}) and (\ref{sigdef2}) using a  measurement rate 
\beq
	\bar{M}  >  2 \lim_{\Delta \rightarrow 0} R_F[\mathcal{X}(\Delta)].
\label{reducesusual}
\eeq
\end{theorem}
%
%
\begin{theorem}
\label{thm4}
\emph{(Converse)}. In the absence of measurement error,
we cannot perfectly recover all signals $f$ satisfying (\ref{sigdef1}) and (\ref{sigdef2})
using a  measurements rate
\beq
	\bar{M}  \leq 2 \lim_{\Delta \rightarrow 0}  R_F[\mathcal{X}(\Delta)].
	\label{reduceusual1}
\eeq
\end{theorem}
In   section VI, we also show that
\beq
\label{equifinal1}
	R_F[\mathcal{X}(\Delta) \oplus \mathcal{X}(\Delta)] = 2 R_F[\mathcal{X}(\Delta)],
	\eeq
which also  implies
\beq
	\lim_{T \rightarrow \infty} \frac{\mbox{dim}_F[\mathcal{X}(\Delta)  \oplus \mathcal{X}(\Delta)]}{  \mbox{dim}_F[\mathcal{X}(\Delta)]} = 2.
\eeq

%
%

We now give a  geometric interpretation of these results.
The set of all multi-band signals is  the union of infinitely many subsets, each corresponding to the multi-band signals of  a given sub-band allocation.  The Minkowski sum in (\ref{eq:minsum})  takes into account the additional degrees of freedom of allocating the sub-bands in the frequency domain. Within any subset,
any multi-band signal is specified   by essentially $\mbox{dim}_F[\mathcal{X}(\Delta)]$ coordinates,  but when considering  the union of all subsets,  
it is specified by essentially $2 \mbox{dim}_F[\mathcal{X}(\Delta)]$ coordinates. By \eqref{equifinal1} it then follows that the relevant information-theoretic quantity that characterizes the possibility of reconstruction is the fractal dimension rate of the dilation, rather than the fractal dimension rate of the set itself.

Finally, it is useful introduce the \emph{sparsity fraction}   as the ratio  of the fractal dimension of the approximating set and its ambient dimension:
\begin{definition} \emph{(Sparsity fraction).} \label{defsparsity}
\beq
\sigma =  \inf_{\nu>0} \lim_{\Delta \rightarrow 0} \lim_{T \rightarrow \infty}  \frac{\text{\emph{dim}}_F[\mathcal{X}(\Delta)]}{N}.
\label{defsparsityeq}
\eeq
\end{definition}
By the results above, it is   easy to see that the sparsity fraction is equal to the fraction of occupied bandwidth, namely substituting $N=(1+\nu)\Omega T/\pi$ into \eqref{defsparsityeq} we get
\beq
\sigma =  \inf_{\nu>0} \lim_{\Delta \rightarrow 0} \frac{R_F[\mathcal{X}(\Delta)]  }{\Omega  } \frac{\pi}{(1+\nu)}
=\frac{\Omega'}{\Omega},
\eeq
and twice the sparsity fraction   corresponds to the critical number of measurements  per unit ambient dimension necessary and sufficient for reconstruction.

\subsection{General Case}
 Results generalize to the noisy case. The critical threshold for the number of measurements is not affected by the presence of a measurement error, provided that we ask for robust, rather than perfect reconstruction.
\begin{theorem}
\label{thm5} 
\emph{(Direct)}. We can robustly recover all signals $f$ satisfying (\ref{sigdef1}) and (\ref{sigdef2})	using a measurements rate
\beq
	\bar{M}  >  2 \lim_{\Delta \rightarrow 0} R_F[\mathcal{X}(\Delta)] =\frac{2 \Omega'}{\pi}.
\eeq
\end{theorem}
\begin{theorem}
\label{thm6} 
\emph{(Converse)}. We cannot robustly recover all signals $f$ satisfying (\ref{sigdef1}) and (\ref{sigdef2})	using a  measurements rate
\beq
	\bar{M}  \leq 2 \lim_{\Delta \rightarrow 0}  R_F[\mathcal{X}(\Delta)] =\frac{2 \Omega'}{\pi}.
\eeq
\end{theorem}
A factor of two is the price to pay for blindness for both robust recovery and perfect recovery of multi-band signals, and in virtue of \eqref{equifinal1} the relevant dimensionality notion is the one associated to the dilation of the set. 

\section{Comparison with compressed sensing}
There are analogies between our results and the ones in compressed sensing. We illustrate similarities and differences in   deterministic and   stochastic settings. For  simplicity, we only consider the case of zero  measurement error, but the same considerations apply to the case of non-zero measurement error.

\label{sec:result}

 \subsection{Deterministic setting}
 
\begin{figure}[!t]
	\centering
	\includegraphics[width=90mm]{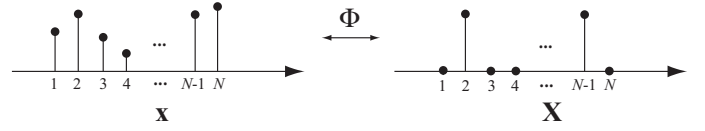}
	\caption{Illustration of a discrete vector with a sparse representation.}
	\label{fig1}
\end{figure}

Consider an $N$-dimensional vector $\bf{x}$ such that
\beq
\label{eq:cs}
	{\bf{x}} = \Phi {\bf{X}},
\eeq
where $\Phi$ is an $N \times N$ orthogonal matrix 
and  $\bf{X}$ has at most $S$ non-zero elements. If $S \ll N$ we say that     $\mathbf{X}$ is a sparse representation of   ${\bf x}$.   An example is illustrated in Figure~\ref{fig1}.

We define a measurement vector 
\beq
	{\bf{y}} =  {\rm{A }}{\bf{x}},
	\label{eq:csm}
\eeq
where $\rm{A}$ is an $M \times N$ matrix, and $M$ is   the number of measurements.  
Cleary, $\bf{x}$ can be recovered from $N$ measurements by observing all the elements of $\bf{x}$.  In this case, the $N \times N$ measurement matrix $\rm{A}$ is diagonal.  If we know the position of the nonzero elements of $\textbf{X}$, then $S$ measurements are also enough to perfectly reconstruct ${\bf x}$. 
In this case, each measurement extracts the $n$th coefficient of  $\textbf{X}$ from  $\Phi^{-1} {\bf x}$, and the signal is recovered by  performing a final multiplication by $\Phi$. 
 However, if  we only know that $\bf{x}$  has a sparse representation, but we do not know the positions of the nonzero elements of ${\bf X}$, without further investigation we can only conclude that that the minimum number $M$ of measurements sufficient for reconstruction is $S \leq M \leq N$. The objective of compressed sensing is to reconstruct any sparse, discrete signal ${\bf x}$  using $M \ll N$ measurements~\cite{csbook}.

Without worrying about an explicit reconstruction procedure, a simple linear algebra argument \cite[Remark 2]{Wu2010}, \cite[Section 2.2]{csbook} shows that the necessary and sufficient number of measurements for reconstruction is $2S$. It follows that in both the continuous and discrete settings, the number of linear measurements necessary and sufficient for reconstruction is equal to twice the sparsity level of the signal.
The main differences between the two settings are as follows: the compressed sensing formulation assumes knowledge of the matrix $\Phi$, corresponding to the basis where the discrete signal is sparse. In the case of blind sensing, it is only assumed that the signal does not occupy the whole frequency spectrum, but the discrete basis set required for the optimal representation is unknown a priori. A   more extreme situation is the blind compressed sensing set-up~\cite{Gleichman2011, Bres}, where there is a complete lack of knowledge about the signal. In this case,    the basis must either be learned from data, or selected from a restricted set. Finally, in blind sensing the reconstruction error tends to zero as $T \rightarrow \infty$, while in compressed sensing perfect reconstruction is possible for all $N$.


 \subsection{Stochastic setting}  
 
\begin{figure}
\begin{center}
{
{\includegraphics[width=85mm]{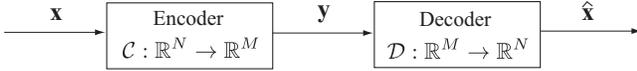}}}
\end{center}
\caption{Source coding view of compressed sensing.
}
\label{csensing}
\end{figure}

The problem of compressed sensing can also be formulated in a probabilistic setting.  In this case,  
the discrete signal to be recovered is modeled as a stochastic process and the objective is to reconstruct the signal with arbitrarily small probability of error, given a sufficiently long observation sequence.  Viewing the measurement operator as an encoder and the reconstruction operator as a decoder acting on a sequence of independent, identically distributed (i.i.d.), real-valued random variables, the compressed sensing set-up corresponds to lossless source coding \index{source coding} of analog memoryless sources when the  encoding operation $\mathcal{C}: \mathbb{R}^N \rightarrow \mathbb{R}^M$ is the multiplication by a real-valued matrix, see Figure~\ref{csensing}. 
Compared to the deterministic setting, where   reconstruction is required  for all possible source signals,  here the performance is measured on a probabilistic basis by considering long block lengths and averaging with respect to the distribution of the source signal.  Compared to the continuous  setting, probabilistic concentration is used to bound the error performance as $N \rightarrow \infty$, instead than spectral concentration as $T \rightarrow \infty$.

Modeling ${\bf x}$  in (\ref{eq:csm}) as a random vector   composed of $N$  independent random variables $(\mathsf{X}_1,\mathsf{X}_2, \ldots \mathsf{X}_N)$, all
distributed as $\mathsf{X}$, 
to capture the notion of sparsity in a probabilistic setting    we may consider the following mixture distribution for the source sequence  
\beq
p_{\mathsf{X}}(x) = (1-\gamma) \delta(x) + \gamma p'(x),
\label{eq:spargamma}
\eeq
where $\delta(\cdot)$ is Dirac's distribution, $0 \leq \gamma \leq 1$, and $p'$ is an absolutely continuous probability measure~\footnote{Results hold more generally for  discrete-continuous mixtures, not only when the discrete part is a  Dirac's distribution.}. 

By the law of large numbers, the parameter $\gamma$ in (\ref{eq:spargamma})    represents, for large values of $N$, the level of sparsity of the signal in terms of  the fraction of its nonzero elements. Given this source model, a basic result  for probabilistic reconstruction by Wu and Verd\'{u}  \cite{Wu2010, Wu2012} shows  that the threshold for 
  the smallest measurement rate that guarantees reconstruction with vanishing probability of error  is independent of  the prior distribution of the non-zero elements $p'$, and equals the sparsity level $\gamma$. 
Comparing this result with the   deterministic case, it follows  that  probabilistic reconstruction, rather than reconstruction for all signals in the space,  yields an improvement in the number of measurements of a factor of two. 

Wu and Verd\'{u} also showed that their result can   be viewed in terms of the infomation (R\'{e}nyi) dimension of the source process. This is somewhat analogous to a coding theorem, where an operational quantity, such as the smallest  rate for recovery, is shown to be equal to an information-theoretic one.
Consider the quantized version  $\mathsf{X}^\epsilon$ of $\mathsf{X}$ obtained from the discrete probability measure induced by partitioning the real line into intervals of size $\epsilon$ and assigning to the quantized variable the probability of lying in each interval. 
The R\'{e}nyi dimension of $\mathsf{X}$ is  defined as~\cite{Reny1}
\begin{definition} \emph{(Information dimension).}
\beq
\mbox{\emph{dim}}_I(\mathsf{X}) =  \lim_{\epsilon \rightarrow 0}  \frac{H_{\mathsf{X}^{\epsilon}}}{- \log \epsilon},
\label{eq:defreninfo}
\eeq
where $H_{\mathsf{X}^{\epsilon}}$ indicates the Shannon entropy of $\mathsf{X}^{\epsilon}$.
\end{definition}
In the case the limit in (\ref{eq:defreninfo}) does not exist, then lower and upper information dimensions  are defined by taking lim inf and lim sup, respectively. 

The definition  immediately extends to a sequence of $N$ i.i.d.  random variables
\beq
\mbox{dim}_I(\mathsf{X_1}, \mathsf{X_2}, \ldots \mathsf{X}_N) = N \mbox{dim}_I(\mathsf{X}),
\label{ngamma}
\eeq
and should be compared with Definition~\ref{fractaldimension}   for continuous signals in a deterministic setting.

We can also give an information-theoretic definition of the sparsity fraction in the stochastic setting that is analogous to Definition~\ref{defsparsity}.
\begin{definition} \emph{(Sparsity fraction ---stochastic setting).}
\beq
\gamma = \frac{\mbox{\emph{dim}}_I(\mathsf{X_1}, \mathsf{X_2}, \ldots \mathsf{X}_N)}{N}. 
\eeq
\end{definition}
For a mixture distribution such as (\ref{eq:spargamma}), assuming $H(\lfloor \mathsf{X} \rfloor)<\infty$, R\'{e}nyi showed that~\cite{Reny1}
\beq
\mbox{dim}_I(\mathsf{X}) = \gamma.
\label{renicomb}
\eeq
Combining this result with (\ref{ngamma}) it follows  that the sparsity fraction is also equal to $\gamma$, and the  fraction of non-zero elements of the signal coincides with the information dimension per unit ambient dimension.  In the analogous   deterministic setting,   the    fraction of occupied bandwidth plays the role of the fraction of non-zero elements of the discrete-time signal, and this coincides with the fractal dimension per unit ambient dimension of its prolate spheroidal approximation.

\subsection{Coding theorems}
The results  of Wu and Verd\'{u} combined with R\'{e}nyi's  one in  (\ref{renicomb}) yield the following general coding theorem:
\begin{theorem} \emph{(Coding theorem ---stochastic setting)}. \\
 The minimum number of measurement per unit dimension sufficient for   reconstruction  with vanishing probability of error  of  an analog, $\gamma$-sparse, memoryless, discrete-time process coincides with  the information dimension per unit ambient dimension of the space, which is equal to $\gamma$. 
 \end{theorem}
 The analogous  deterministic coding theorem   in our continuous  setting is obtained by combining Theorems~\ref{thm3} and \ref{thm4}, and using   Definition~\ref{defsparsity}:  
 \begin{theorem} \emph{(Coding theorem ---deterministic setting)}. \\
The minimum number of measurement per unit dimension sufficient for reconstruction with vanishing error  of any $\sigma$-sparse, continuous-time signal  coincides with twice the fractal dimension per unit ambient dimension of   its   prolate spheroidal   approximation, which is equal to $2 \sigma$. 
\end{theorem}
A factor of two appears in the deterministic formulation, due to the worst case reconstruction scenario.

\section{Technical Preliminaries}
\label{sec:prelim}

\subsection{Metric spaces}
We begin our proofs by defining the metric spaces associated to the bandlimited and multi-band signals satisfying (\ref{sigdef1}) and (\ref{sigdef2}).
Let $f \in L^2(-\infty,\infty)$,   $2 \Omega' < \Omega$, and
\begin{align}
	\BO &= \{ f(t) : \mathfrak{F}f(\omega)  = 0, \mbox{ for } |\omega| > \Omega \}, \\
	\mathcal{B}_{\mathcal{Q}} &= \{ f(t) : \mathfrak{F}f(\omega)  = 0, \mbox{ for } \omega \notin 
	\mathcal{Q} \}, \\
\mathcal{Q'} &=\{ \mathcal{Q} : \mathcal{Q} \subset [-\Omega, \Omega] {\rm{~and~}} m(\mathcal{Q}) \leq 2\Omega' \}, \\
\BQ &= \bigcup_{\mathcal{Q} \in \mathcal{Q'}} \mathcal{B}_{\mathcal{Q}}. \label{eq:BQ}
\end{align}
 It follows that $\BQ \subset \BO$.
We equip $\BO$ and $\BQ$ with the $L^2[-T/2,T/2]$ norm
\beq
\label{eq:normdef1}
	\|f\|= \left( \int_{-T/2}^{T/2} f^2(t) dt \right)^{1/2}.
\eeq
It follows that $(\BO, \|\cdot\|)$ and $(\BQ, \|\cdot\|)$ are metric spaces, whose elements are square-integrable,
real, bandlimited or multi-band signals, of infinite duration and observed over the finite interval $[-T/2,T/2]$.

\subsection{Optimal  representations}

Let $\mathcal{Q} $ be a measurable subset of $\mathbb{R}$ and $\mathcal{T}=[-T/2,T/2]$.
We define the following   time-limiting and band-limiting operators
\begin{align}
	\mathfrak{T}_{\mathcal{T}} f(t) &= \mathbbm{1}_{\mathcal{T}} f(t)
	\\
	\mathfrak{B}_{\mathcal{Q}} f (t) &= \mathfrak{F}^{-1} \mathbbm{1}_{\mathcal{Q}} \mathfrak{F} f(t),
\end{align}
where $\mathbbm{1_{(\mathcal{\cdot)}}}$ is the indicator function. We consider the following eigenvalues equation  
\beq
\label{eq:integral}
	\mathfrak{T}_{T} \mathfrak{B}_{\mathcal{Q}} \mathfrak{T}_{\mathcal{T}} \psi^{\mathcal{Q}} (t) = \lambda^{\mathcal{Q}} \psi^{\mathcal{Q}}(t).
\eeq

There exists a countably infinite set of real functions $\{ \psi_n^{\mathcal{Q}}(t) \} _{n=1} ^{\infty}$ and a set of real positive numbers $1>\lambda_1^{\mathcal{Q}}>\lambda_2^{\mathcal{Q}}>\cdots >0$ with the following properties, see~\cite{Landau1985}.

{\emph{Property 1.}}
The elements of $\{\lambda_n^{\mathcal{Q}}\}$ and $\{\psi_n^{\mathcal{Q}} (t) \}$ are solutions of  (\ref{eq:integral}).

{\emph{Property 2.}}
The elements of $\{\psi_n^{\mathcal{Q}}(t)\}$ are in $\mathcal{B}_\mathcal{Q}$.

{\emph{Property 3.}}
$\{ \psi_n^{\mathcal{Q}}(t) \} $ is complete in $\mathcal{B}_{\mathcal{Q}}$. 

{\emph{Property 4.}}
The elements of $\{ \psi_n^{\mathcal{Q}}(t) \}$  are orthonormal in   $(-\infty, \infty)$.

{\emph{Property 5.}}
The elements of $\{ \psi_n^{\mathcal{Q}}(t) \}$ are orthogonal in  $\left(-T/2, T/2 \right)$ 
\beq
\label{p5}
	\int_{-T/2}^{T/2} \psi_n^{\mathcal{Q}}(t) \psi_m^{\mathcal{Q}}(t) dt = 
		\begin{cases}
		\lambda_n^{\mathcal{Q}} & n=m,\\
		0 & \mbox{otherwise}.
		\end{cases}
\eeq

We write $\psi(t) \mbox{ and } \lambda$ instead of $\psi^{\mathcal{Q}}(t) \mbox{ and } \lambda^{\mathcal{Q}}$ when $\mathcal{Q}=[-\Omega, \Omega]$. In this special case, the eigenfunctions  $\{\psi_n(t)\}$ are the prolate spheroidal wave functions (PWSF)~\cite{Flammer}.
\begin{lemma} \label{lslepia}\emph{(Slepian \cite{SlepianS1}).}
For any $\nu>0$, $N = (1+\nu) \Omega T/\pi$, and $f \in \BO$, there exist real coefficients $\{x_n\}$, such that the approximation
\beq
\label{expansion2}
	f_N(t) = \sum_{n=1}^{N} x_{n} \psi_n (t)
\eeq
has vanishing error norm  $\| f- f_N\|$,  as $T \rightarrow \infty$.
\end{lemma}

\begin{lemma} \label{llanda} \emph{(Landau and Widom~\cite{Landau1980}).}
 For any $\nu>0, S = (1+\nu) \Omega' T/\pi$,  and $f \in \BQ$,  there exist real coefficients $\{\alpha_n\}$, such that the  approximation
\beq
\label{expansion1}
	f_S(t) = \sum_{n=1}^{S} \alpha_{n} \psi_n^{\mathcal{Q}} (t),
\eeq
has vanishing error norm $\| f- f_S\|$, as $T \rightarrow \infty$.
\end{lemma}

\subsection{Measurement vector}
We consider the  measurements of  $f(t) \in  \BQ \subset \BO$
\begin{align}
	y_n & = \int_{-T/2}^{T/2} f(t) \varphi_n(t) dt  + e_n, \;\;\;   n \in \{1,\ldots, M\},
\end{align} 
where $e_n$ is the measurement error and each measurement kernel $\varphi_n$ is a bandlimited function. Since $\varphi_n$ is bandlimited, this can be represented by a linear combination of the ``canonical'' PSWF basis of $\BO$, namely
\beq
	\varphi_n(t) = \sum_{k=1}^{\infty} a_{nk} \psi_k (t)
\eeq
Using the completeness    of the $\{\psi_n(t) \}$ in $\BO$, and their orthogonality property, it follows that the $n$-th measurement    can also be expressed as  
\begin{align}
	y_n & = \int_{-T/2}^{T/2} f(t) \varphi_n(t) dt + e_n \nonumber
	\\
	& = \int_{-T/2}^{T/2} {\sum_{j=1}^{\infty} x_j \psi_j (t)} {\sum_{k=1}^{\infty} a_{nk} \psi_k (t)} dt + e_n \nonumber
	\\
	& = \sum_{j=1}^{N} a_{nj} x_j \sqrt{\lambda_j} + \sum_{j=N+1}^{\infty} a_{nj} x_j \sqrt{\lambda_j} + e_n.
\end{align}
Letting $N=(1+\nu) \Omega T/\pi$, we have
\beq
\lim_{T \rightarrow \infty} \sum_{j=N+1}^{\infty} a_{nj} x_j \sqrt{\lambda_j} =0.
\label{virtue}
\eeq 
 It  follows that as $T \rightarrow \infty$ the  measurements become
\beq
\label{eq:measure}
	y_n = \sum_{j=1}^{N} a_{nj} x_j \sqrt{\lambda_j} + e_n + o(1).
\eeq
Letting ${\bf{y}}=(y_1, \cdots, y_M)$, ${\bf{x}}=(x_1 \sqrt{\lambda_1}, \cdots, x_N \sqrt{\lambda_N})$, and $A$ be an $M \times N$ matrix  such that $[A]_{nj} = a_{nj}$, we   define 
\beq
\label{eq:measurement}
	{\bf{y}} = {\rm{A}} {\bf{x}} + {\rm{e}},
\eeq
%
%
and  consider the  set
\beq
	\overline{\mathcal{X}} = \left\{ {\bf{x}} : {\bf{x}} = \left(x_1 \sqrt{\lambda_1}, \cdots, x_N \sqrt{\lambda_N} \right) \right\}.
		\label{xbar}
\eeq
In virtue of Lemma~\ref{lslepia}, there exists a  one-to-one correspondence between $\BQ$ and $\overline{\mathcal{X}}$, as $T \rightarrow \infty$. By \eqref{eq:measure} it then follows that to complete our proofs
we can   derive lower and upper bounds  on the number of rows of $\rm{A}$ required to recover ${\bf{x}} \in \overline{\mathcal{X}}$ from   ${\bf{y}} = {\rm{A}} {\bf{x}} + {\rm{e}}$ in (\ref{eq:measurement}), and then evaluate their   order of growth as $T \rightarrow \infty$.   


\section{Proofs of Theorems \ref{thm1} and  \ref{thm2}}
\label{sec:proofs1}

We consider a function $\zeta(t)$ such that
\beq
	f_N(t) = f_S (t) + \zeta(t),
\eeq
where $f_N(t)$ and $f_S(t)$ are given in (\ref{expansion2}) and (\ref{expansion1}), and let
\beq
	\zeta_k = \int_{-T/2}^{T/2} \zeta(t) \psi_k (t) dt.
\eeq 
It follows that for all  $1 \leq k \leq N$, we have
\begin{align}
	\sqrt{\lambda_k} x_k & = \int_{-T/2}^{T/2} \sum_{n=1}^{S} \alpha_{n} \psi_n^{\mathcal{Q}} (t) \psi_k (t) dt+ \zeta_k \nonumber
	\\
	&= \sum_{n=1}^{S} \alpha_n \int_{-T/2}^{T/2} \psi_n^{\mathcal{Q}} (t) \psi_k (t) dt +\zeta_k .	\
\end{align}
We now define 
\beq
\varphi_{k,n}^{\mathcal{Q}} = \int_{-T/2}^{T/2} \psi_n^{\mathcal{Q}} (t) \psi_k (t) dt,
\label{quaranta}
\eeq 
so that we have
\begin{align}
\sqrt{\lambda_k} x_k 	&= \sum_{n=1}^{S} \alpha_n \varphi_{k,n}^{\mathcal{Q}} +\zeta_k.
\label{tovector}
\end{align}
We rewrite (\ref{tovector}) in vector form as
\beq
\label{exact}
	{\bf{x}} = \Phi_{\mathcal{Q}} {\bf{\alpha}} + {\bf{\zeta}},
\eeq
where ${\bf x } \in \overline{\mathcal{X}}$,   
${\bf{\alpha}}=(\alpha_1, \cdots, \alpha_S)$,   ${\bf{\zeta}}=(\zeta_1 \cdots, \zeta_N)$,     and $\Phi_{\mathcal{Q}}$  is an $N \times S$ matrix such that 
\beq
 [\Phi_{\mathcal{Q}}]_{k,n} = \varphi_{k n}^{\mathcal{Q}}.
 \label{matrvec}
\eeq
By Lemmas~\ref{lslepia} and \ref{llanda}, we have that $\bf{\zeta}$ tends to the all zero vector as $T \rightarrow \infty$. Therefore, it is enough to  determine the minimum number of measurements to recover
\beq
\label{approx}
	{\bf{x}} = \Phi_{\mathcal{Q}} {\bf{\alpha}}.
\eeq

Let us define the following set
\beq
	\mathcal{D} = \{ \Phi_{\mathcal{Q}} : \mathcal{Q} \in \mathcal{Q'} \}.
\eeq
We  rewrite $\overline{\mathcal{X}}$ in (\ref{xbar}) as follows:
\beq
	\overline{\mathcal{X}}= \{ {\bf{x}} : {\bf{x}} = \Phi_{\mathcal{Q}} {\bf{\alpha}} {\rm{~where~}} \Phi_{\mathcal{Q}} \in \mathcal{D} {\rm{~and~}} {\bf{\alpha}} \in \mathbb{R}^{S} \}.
\eeq

\begin{lemma}
\label{lemmaA1}
For all $\Phi_1, \Phi_2 \in \mathcal{D}$,
	there exists an $m \times N$ matrix $\rm{A}$ such that ${\rm{rank}}(A[\Phi_1,\Phi_2]) = {\rm{rank}}[\Phi_1,\Phi_2]$, provided that
	\beq
		m \geq \max_{\Phi_1, \Phi_2 \in \mathcal{D}} \left( {\rm{rank}} [ \Phi_1 , \Phi_2 ] \right).
	\eeq
\end{lemma}
\begin{IEEEproof}
It is enough to show that for all $\Phi_1, \Phi_2 \in \mathcal{D}$, if $\rm{A}$ is an i.i.d Gaussian random matrix of size $m \times N$, then  ${\rm{rank}}({\rm{A}}[\Phi_1,\Phi_2]) = {\rm{rank}}([\Phi_1,\Phi_2])$ with probability 1. 
Since ${\rm{rank}}({\rm{A}}[\Phi_1,\Phi_2]) \leq {\rm{rank}}([\Phi_1,\Phi_2])$, it is enough to show that ${\rm{rank}}({\rm{A}}[\Phi_1,\Phi_2]) \geq {\rm{rank}}([\Phi_1,\Phi_2])$. For  convenience, we let $[\Phi_1 , \Phi_2] = \Phi$ and we will show ${\rm{rank}}({\rm{A}} \Phi) \geq {\rm{rank}}(\Phi)$. 

Note that $\Phi$ is an $N \times 2S$ matrix with ${\rm{rank}}(\Phi) = r \leq m$. 
Collect $r$ independent columns of $\Phi$ and compose an   $N \times r$ matrix $\Phi'$. 
Using the Gram-Schmidt process, we can transform $\Phi'$ into $\Phi_G$, an $N \times r$ matrix, whose columns are orthonormal. 
By adding redundant $N-r$ orthonormal columns followed by the original $r$ columns of $\Phi_G$, we obtain an $N \times N$ orthogonal matrix $\overline{\Phi}_G$. 

Let us define $\sigma(\rm{X})$ as the smallest   number of linearly dependent columns of a matrix $\rm{X}$.  
It is well known that, if $\rm{A}$ is an i.i.d. Gaussian random matrix of size $m \times N$, where $m <N$, then $\sigma({\rm{AP}}) = m+1$ with probability 1 for any fixed orthogonal matrix $\rm{P}$, see for example \cite[Proposition 1]{Gleichman2011} for a   proof.
Therefore, the first $r$ columns of ${\rm{A}} \overline{\Phi}_G$ are independent. Thus, we have ${\rm{rank}} ({\rm{A}} \Phi_G) =r$, which implies ${\rm{rank}}({\rm{A}} \Phi') =r$. We can then conclude that ${\rm{A}}\Phi$ contains at least $r$ independent columns, which implies ${\rm{rank}}({\rm{A}}\Phi) \geq r = {\rm{rank}}(\Phi)$. 

\end{IEEEproof}

\begin{lemma}
\label{lemmaA2}
	A   number of measurements 
	\beq
		m \geq \max_{\Phi_1, \Phi_2 \in \mathcal{D}} \left( {\rm{rank}} [ \Phi_1 , \Phi_2 ] \right),
	\eeq
	is sufficient to recover all the elements of $\overline{\mathcal{X}}$.
\end{lemma}
\begin{IEEEproof}
From Lemma \ref{lemmaA1} it follows that for all $\Phi_1, \Phi_2 \in \mathcal{D}$ there exists an $m \times N$ matrix $\rm{A}$ such that ${\rm{rank}}({\rm{A}} [\Phi_1,\Phi_2]) = {\rm{rank}}[\Phi_1,\Phi_2]$. 
Let us assume ${\rm{A}}{\bf{x_1}} = {\rm{A}}{\bf{x_2}}$ where ${\bf{x_1}} = \Phi_1 \alpha_1$ and ${\bf{x_2}} = \Phi_2 \alpha_2$. 
The expression ${\rm{A}}{\bf{x_1}} = {\rm{A}}{\bf{x_2}}$, can be rewritten as
\begin{align}
	{\rm{A}} [\Phi_1 , \Phi_2] 
	\begin{bmatrix}
		\alpha_1 \\ \alpha_2
	\end{bmatrix}
	=0,
\end{align}
namely $[\alpha_1 , \alpha_2]^T$ belongs to the null space of ${\rm{A}}[\Phi_1 , \Phi_2 ]$.
Since ${\rm{rank}}({\rm{A}}[\Phi_1,\Phi_2]) = {\rm{rank}}[\Phi_1,\Phi_2]$, the null space of ${\rm{A}}[\Phi_1,\Phi_2]$ is the same as the null space of $[\Phi_1,\Phi_2]$. 
It follows that $[\alpha_1 , \alpha_2]^T$ belongs to the null space of $[\Phi_1 , \Phi_2 ]$, or equivalently
\begin{align}
	[\Phi_1 , \Phi_2] 
	\begin{bmatrix}
		\alpha_1 \\ \alpha_2
	\end{bmatrix}
	=0.
\end{align}
This means $\Phi_1 \alpha_1  = \Phi_2 \alpha_2$, namely ${\bf{x_1}} = {\bf{x_2}}$. Therefore, ${\rm{A}}$ is one-to-one on $\overline{\mathcal{X}}$, which implies that the elements of $\overline{\mathcal{X}}$ can be recovered.

\end{IEEEproof}


\begin{lemma}
\label{lemmaA3}
	A number of measurements 
	\beq
		m < \max_{\Phi_1, \Phi_2 \in \mathcal{D}} \left( {\rm{rank}} [ \Phi_1 , \Phi_2 ] \right)
	\eeq
	is not sufficient to recover all the elements of $\overline{\mathcal{X}}$.
	
\end{lemma}
\begin{IEEEproof}
If all elements ${\bf{x}} \in \overline{\mathcal{X}}$ can be recovered from the measurements ${\bf{y}} = {\rm{A}} {\bf{x}}$, where ${\rm{A}}$ is an $m \times N$ matrix, this means ${\rm{A}}$ is one-to-one on $\overline{\mathcal{X}}$. 
Therefore, for all ${\bf{x_1}}$ and ${\bf{x_2}}$, ${\rm{A}}{\bf{x_1}} = {\rm{A}}{\bf{x_2}}$   implies ${\bf{x_1}} = {\bf{x_2}}$. 
Let us assume ${\bf{x_1}} = \Phi_1 \alpha_1$ and ${\bf{x_2}} = \Phi_2 \alpha_2$, then ${\rm{A}} \Phi_1 \alpha_1 = {\rm{A}} \Phi_2 \alpha_2$   implies  $\Phi_1 \alpha_1 = \Phi_2 \alpha_2$. 
This is equivalent to saying that
\begin{align}
	{\rm{A}} [\Phi_1 , \Phi_2] 
	\begin{bmatrix}
		\alpha_1 \\ \alpha_2
	\end{bmatrix}
	=0
\end{align}
implies
\begin{align}
	[\Phi_1 , \Phi_2] 
	\begin{bmatrix}
		\alpha_1 \\ \alpha_2
	\end{bmatrix}
	=0.
\end{align}
Namely, the null space of ${\rm{A}} [\Phi_1 \Phi_2]$ is contained in the null space of $[\Phi_1 \Phi_2]$. By the rank-nullity theorem, we have
\beq
\label{temp1-1}
	{\rm{rank}} ({\rm{A}} [\Phi_1 , \Phi_2]) \geq {\rm{rank}} [\Phi_1 , \Phi_2].
\eeq
Since $m \geq {\rm{rank}} ({\rm{A}} [\Phi_1 , \Phi_2])$ and (\ref{temp1-1})  holds for all $\Phi_1, \Phi_2 \in \mathcal{D}$,   the  result follows. 

\end{IEEEproof}

\begin{lemma}
\label{lemmaA4}
	We have 
	 
	\beq
	\lim_{T\rightarrow \infty} \frac{	\max_{\Phi_1, \Phi_2 \in \mathcal{D}} \left( {\rm{rank}} [ \Phi_1 , \Phi_2 ] \right)}{2S}=1.
	\eeq
	
	\end{lemma}
\begin{IEEEproof}
Let   $\mathcal{Q}_1, \mathcal{Q}_2 \in \mathcal{Q'}$,  and
consider the multi-band signals
\beq
	f_1(t) = \sum_{n=1}^{S} \alpha_n \psi_n^{{\mathcal{Q}}_1} (t), 
	\label{effe1}
\eeq
\beq
	f_2(t) = \sum_{n=1}^{S} \beta_n \psi_n^{{\mathcal{Q}}_2} (t), 
	\label{effe2}
\eeq
and 
\beq
f_S(t)= f_1(t) + f_2(t).
\label{plusf1f2}
\eeq
Consider the  $N$-dimensional vector 
\begin{align}
 {\bf{z}} =
	[\Phi_{\mathcal{Q}_1} , \Phi_{\mathcal{Q}_2}] 
	\begin{bmatrix}
		\alpha_1 \\ \vdots \\ \alpha_S \\ \beta_1 \\ \vdots \\ \beta_S,
	\end{bmatrix}.
\end{align}
whose elements, by   (\ref{quaranta}) and (\ref{matrvec}), and then using (\ref{effe1}), (\ref{effe2}),   (\ref{plusf1f2}), are
\begin{align}
	z_n & = \sum_{j=1}^{S} \alpha_j \varphi_{n,j}^{\mathcal{Q}_1} + \sum_{j=1}^{S} \beta_j \varphi_{n,j}^{\mathcal{Q}_2} \nonumber \\
	&=  \sum_{j=1}^{S} \alpha_j  \int_{-T/2}^{T/2} \psi_j^{{\mathcal{Q}}_1} (t)  \psi_n (t) dt \nonumber \\
	& + 	\sum_{j=1}^{S} \beta_j \int_{-T/2}^{T/2}   \psi_j^{{\mathcal{Q}}_2} (t) \psi_n (t) dt \nonumber \\
	& = 	   \int_{-T/2}^{T/2} f_1(t) \psi_n(t) dt +   \int_{-T/2}^{T/2} f_2(t) \psi_n(t) dt \nonumber \\
	& = \int_{-T/2}^{T/2} f_S(t) \psi_n (t) dt.  \label{sco}
\end{align}


We   consider the case when ${\bf{z}}$ is the all zero vector. In this case, since  by (\ref{sco})  the elements $\{ z_n\}$ are also the PSWF coefficients of $f_S(t)$, it follows that 
\beq
	\lim_{T \rightarrow \infty} f_S(t) =0.
	\label{middle}
\eeq
We now choose ${\mathcal{Q}}_1$ and ${\mathcal{Q}}_2$ such that ${\mathcal{Q}}_1 \cap {\mathcal{Q}}_2 = \emptyset$, so that (\ref{middle}) implies
\beq
	\lim_{T \rightarrow \infty} f_1(t) = \lim_{T \rightarrow \infty} f_2(t) = 0.
\eeq
It follows that all coefficients $\{ \alpha_n \}$ and $\{ \beta_n \}$ in (\ref{effe1}) and (\ref{effe2})   must tends to zero as $T \rightarrow \infty$,    
 the columns of $[\Phi_{\mathcal{Q}_1} , \Phi_{\mathcal{Q}_2}]$ become independent, and we have
 \beq
 \lim_{T \rightarrow \infty} \frac{ {\rm{rank}}[\Phi_{\mathcal{Q}_1}, \Phi_{\mathcal{Q}_2}] }{2 S} =1.
 \eeq
On the other hand, ${\rm{rank}}[\Phi_1, \Phi_2] \leq 2S$ for all  $\Phi_1, \Phi_2 \in \mathcal{D}$ because the number of columns of $[\Phi_1, \Phi_2]$ is $2S$. It follows that our choice   $\Phi_1 = \Phi_{\mathcal{Q}_1}$ and $\Phi_2= \Phi_{\mathcal{Q}_2}$ achieves the maximum rank and  the result   follows.


\end{IEEEproof}

By combining    Lemmas {\ref{lemmaA2}} and   {\ref{lemmaA4}} it follows that with  
$2 S =  2(1+\nu)\Omega'T/\pi$ measurements we can recover any  vector  ${\bf x}$ in (\ref{approx}) with vanishing error as $T \rightarrow \infty$, and since   the vector $\zeta$ tends  zero    we can also  recover any vector {\bf x} in (\ref{exact}). It follows that we can recover  the  coefficients  representing any signal in $\BQ$ with vanishing error using  a measurement rate  
\beq
\bar{M} = \frac{2 \Omega'}{\pi} + 2 \nu \frac{\Omega'}{\pi} > \frac{2 \Omega'}{\pi},
\eeq
and the proof of Theorem~\ref{thm1} is complete.

On the other hand,
by combining Lemmas {\ref{lemmaA3}} and   {\ref{lemmaA4}} it follows that with   less than $2S = 2(1+\nu)\Omega'T/\pi$ measurements we cannot recover all possible vectors ${\bf x}$ in (\ref{approx}) with vanishing error as $T \rightarrow \infty$. This also means that we cannot recover all possible vectors ${\bf x}$ in (\ref{exact}). It follows that with a number of measurements  $M = 2\Omega'T/\pi + o(T),
$
and hence a  measurement rate
\beq
 \bar{M}  = 2 \Omega'/\pi
\eeq
we cannot recover all signals in $\BQ$, and the proof of Theorem~\ref{thm2} is also complete.



\section{Proofs of Theorems \ref{thm3}-\ref{thm6}}
\label{sec:proofs2}

In the following, we use  $\| \cdot \|$ to denote the Euclidean  norm for vectors in $\mathbb{R}^N$
\beq
\| {\bf x} \| =  \sqrt{\sum_{n=1}^N x^2_n},
\eeq
and the spectral norm for matrices
\beq
\| \rm{A} \| = \sup_{\bf x \not = 0} \frac{\| \rm{A} {\bf x} \|}{\| \bf{x}\|}.
\eeq
We also use the usual notation for signals defined in \eqref{eq:normdef1}.

\subsection{The key lemmas}

Let $\mathcal{B}_{\Delta}$  be the collection of all elements in $\BQ$  such that      the extremal points of all sub-bands belong to  the discrete set $\mathcal{J}$ defined in \eqref{mathD}. For any signal $f \in \BQ$,  let $f_{\Delta} \in \mathcal{B}_{\Delta}$ such that 
\beq
	f_{\Delta} = \argmin_{f' \in \mathcal{B}_{\Delta}} \| f - f' \|.
\eeq
Since all $f \in \BQ$ are square-integrable, it follows that
\beq
	\lim_{\Delta \rightarrow 0} \| f - f_{\Delta} \| = 0.
\eeq
Hence,
if $f_{\Delta}$ can be   recovered using a measurement rate $\bar{M}_{\Delta}$, then $f$ can   be   recovered using a measurement rate
\beq
	\bar{M} = \lim_{\Delta \rightarrow 0} \bar{M}_{\Delta}.
\eeq

Consider now the  set  $\overline{\mathcal{X}}(\Delta)$  of vectors  of  $N=(1+\nu) \Omega T/\pi$   real coefficients, such that   every element of  $ \mathcal{B}_{\Delta}$ is approximated, with vanishing error as $T \rightarrow \infty $, by an element of $\overline{\mathcal{X}}(\Delta)$. We also consider $\mathcal{X}(\Delta) \subset \overline{\mathcal{X}}(\Delta)$ containing all elements of $\overline{\mathcal{X}}(\Delta)$ that have norm at most one. 
To prove Theorems \ref{thm3}-\ref{thm6}, it is enough to prove following two lemmas.

\begin{lemma}
\label{lt5} 
We can robustly recover all signals $f \in {\mathcal{B}}_{\Delta}$ 	using a measurements rate
\beq
	\bar{M}_{\Delta}  >   R_F[\mathcal{X}(\Delta) \oplus \mathcal{X}(\Delta)].
\eeq
\end{lemma}
%

\begin{lemma}
\label{lt4}
In the absence of measurement error,
we cannot perfectly recover all signals $f \in {\mathcal{B}}_{\Delta}$ 
using a  measurement rate
\beq
	\bar{M}_{\Delta}  < 2 R_F[\mathcal{X}(\Delta)].
\eeq
\end{lemma}

To see that Theorems \ref{thm3}-\ref{thm6} follow from these two lemmas,  first note that the lemmas imply 
\beq
\label{equi01}
	R_F[\mathcal{X}(\Delta) \oplus \mathcal{X}(\Delta)] \geq 2 R_F[\mathcal{X}(\Delta)],
\eeq
on the other hand, we have
\beq
	\dim_F[\mathcal{X}(\Delta) \oplus \mathcal{X}(\Delta)]
	\leq
	2\dim_F[\mathcal{X}(\Delta)],	
\eeq
which implies
\beq
\label{equi02}
	R_F[\mathcal{X}(\Delta) \oplus \mathcal{X}(\Delta)] \leq 2 R_F[\mathcal{X}(\Delta)].
\eeq
Combining (\ref{equi01}) and (\ref{equi02}) it follows that
\beq
\label{equifinal}
	R_F[\mathcal{X}(\Delta) \oplus \mathcal{X}(\Delta)] = 2 R_F[\mathcal{X}(\Delta)].
\eeq

Theorem \ref{thm5} now follows from Lemma \ref{lt5} and (\ref{equifinal})    by taking the limit for $\Delta \rightarrow 0$, and  Theorem \ref{thm3} follows directly from Theorem \ref{thm5}.
On the other hand,   from  Lemma  \ref{lt4} it follows  by taking the limit for $\Delta \rightarrow 0$ that   with a measurement rate
\beq
	\bar{M}  < \lim_{\Delta \rightarrow 0} 2 R_F[\mathcal{X}(\Delta)] 
	\eeq
  we cannot perfectly recover all signals $f \in \BQ$.  As for the equality, combining this result  with Theorems \ref{thm1}, \ref{thm2}, and \ref{thm3}, we  conclude  that
\beq
\label{B5}
	\lim_{\Delta \rightarrow 0}2 R_F[\mathcal{X}(\Delta)] =   \frac {2 \Omega'}{\pi},
	\eeq 
which completes the proof of Theorem \ref{thm4}.  Theorem \ref{thm6} follows directly from Theorem \ref{thm4}.


\subsection{Proof of Lemma \ref{lt5}}


\begin{definition} \emph{(Inverse Lipschitz condition.)} 
	A matrix  $\rm{A}$ satisfies   the inverse Lipschitz condition on  a set  $\mathcal{U} $  if  
 there exists a constant $ \beta >0$  such that   	for all ${\bf{u}_1, {\bf u}_2} \in {\mathcal{U}}$, we have
	\beq
		\beta \|{ \bf u}_1 - {\bf  u}_2 \| \leq \| \rm{A} {\bf u}_1 - \rm{A} {\bf u}_2\|.
	\eeq
\end{definition}

We claim that  if $\rm{A}$ satisfies   the inverse Lipschitz condition on  $\overline{\mathcal{X}}(\Delta)$, then every ${\bf{x}}\in  \overline{\mathcal{X}}(\Delta)$ can be robustly recovered from ${\bf{y}} = \rm{A}{\bf{x}} + {\bf{e}}$. To prove this claim, consider the following two cases: (a) ${\bf{y}} \in \rm{A} \overline{\mathcal{X}}(\Delta)$, where $\rm{A} \overline{\mathcal{X}}(\Delta)$ is the set $\{ \rm{A} {\bf{x}} : {\bf{x}}  \in \overline{\mathcal{X}}(\Delta) \}$, and (b) ${\bf{y}} \notin \rm{A}  \overline{\mathcal{X}}(\Delta)$.
In the first case, let $\bf{x'}$ be a solution of ${\bf{y}} = \rm{A} {\bf{x'}}$ and let $ \bf{x'}$  be the vector used to recover $\bf{x}$. Then, the recovery error   is bounded as
\beq
	\beta \| {\bf{x}}-{\bf{x'}} \| \leq \|{\rm{A}{\bf{x}} - \rm{A}{\bf{x'}}}\| =  \|\bf{e}\|, 
\eeq
which guarantees robust recovery. On the other hand, if ${\bf{y}} \notin \rm{A}  \overline{\mathcal{X}}(\Delta)$, let ${\bf{x''}} \in \overline{\mathcal{X}}(\Delta)$ such that ${\rm{A}} {\bf{x''}}$ is the closest to $\bf{y}$ among all the elements of ${\rm{A}}  \overline{\mathcal{X}}(\Delta)$. By letting $\bf{x''}$ be the vector used to recover $\bf{x}$, we can bound the recovery error as
\bea
	\beta \| {\bf{x}}-{\bf{x''}} \| 
	&\leq&  \|{\rm{A}{\bf{x}} - \rm{A}{\bf{x''}}}\| 
	\\
	&\leq& \|{\rm{A}}{\bf{x}} -{\bf{y}} \| + \| {\bf{y}} - {\rm{A}}{\bf{x''}} \|
	\\
	&\leq& 2 \| \bf{e} \|,
\eea
which guarantees   robust recovery. The claim now follows and we can proceed to derive a sufficient condition that  ensures   $\rm{A}$ satisfies the inverse Lipschitz condition on the set $\overline{\mathcal{X}}(\Delta)$.

By letting   $\overline{\mathcal{Z}}(\Delta) = \overline{\mathcal{X}}(\Delta) \oplus \overline{\mathcal{X}}(\Delta)$,      the inverse Lipschitz condition is equivalent to stating that for all ${\bf{z}}\in \overline{\mathcal{Z}}(\Delta)$
\beq
\label{rip}
	\beta \|{\bf{z}}\| \leq \|{\rm{A}}{\bf{z}}\|.
\eeq
Consider   the normalized set $\mathcal{Z}^{'}(\Delta) \subset \overline{\mathcal{Z}}(\Delta)$  containing all the elements of $\overline{\mathcal{Z}}(\Delta)$ that are vectors of unit norm, and let $k'=\mbox{\normalfont{dim}}_F[\mathcal{Z}^{'}(\Delta)]$. 
If (\ref{rip}) holds for all ${\bf{z}}\in \overline{\mathcal{Z}}(\Delta)$, then it also holds for all ${\bf{z}}\in \mathcal{Z}^{'}(\Delta)$, and vice versa.
In the following, we   consider $\mathcal{Z}^{'}(\Delta)$ instead than $\overline{\mathcal{Z}}(\Delta)$.

Let $\mathcal{L}_{\epsilon}[\mathcal{Z}^{'}(\Delta)]$ be a minimal $\epsilon$-covering set of $\mathcal{Z}^{'}(\Delta)$, namely a minimum cardinality set such that any point in $\mathcal{Z}^{'}(\Delta)$ is within  distance $\epsilon$ from at least one point of $\mathcal{L}_{\epsilon}[\mathcal{Z}^{'}(\Delta)]$. Let $L_{\epsilon}[\mathcal{Z}{'}(\Delta)] = |\mathcal{L}_{\epsilon}[\mathcal{Z}^{'}(\Delta)] |$.
We need the following preliminary results.

\begin{lemma}\emph{\cite[Fact 2.1.]{Olson2002}}
\label{lemmaB1}
\begin{align}
\mbox{\normalfont{dim}}_F[\mathcal{Z}^{'}(\Delta)] = 	\inf \left\{ d : \forall    \epsilon \in (0,1)  \,   \exists{\gamma>0} :  \right. \nonumber \\
 \left. L_{\epsilon}[\mathcal{Z}^{'}(\Delta)] \leq \gamma \left( \frac{1}{\epsilon} \right)^d \right\}.
\end{align}

\end{lemma}

Let $\mathcal{G}$ be the space of all orthogonal projections in $\mathbb{R}^N$ of rank $m$, and $\mu$ be the invariant measure on $\mathcal{G}$ with respect to orthogonal transformations. 

\begin{definition} \emph{(Shadow of a set)}.  
The shadow of a set $\mathcal{B}$ in $\mathbb{R}^N$ is  
\beq
	S(\mathcal{B}) = \{ {\rm{P}} \in \mathcal{G} :  0 \in {\rm{P}} \mathcal{B} \}.
\eeq	
\end{definition}

\begin{lemma}\emph{\cite[Theorem 5.1.]{Olson2002}}
\label{lemmaB2}
The measure of the shadow of a $\rho$-ball $\mathcal{B}$ centered at a distance $r$ from the origin is bounded as
\beq
	\mu (S(\mathcal{B})) \leq \delta \left( \frac{\rho}{r} \right)^{m},
\eeq
where $\delta$ is a positive constant.
\end{lemma}

We now provide a key lemma.

\begin{lemma}
\label{lemmaB3}
For almost every projection ${\rm{P}}$ of rank $m>k'$, there exists a constant $c$ such that, for all ${\bf{z}} \in \mathcal{Z}^{'}(\Delta)$
\beq
	\|{\rm{P}}{\bf{z}}\| > c \|{\bf{z}}\|.
\eeq

\end{lemma}

\begin{IEEEproof}
From Lemma {\ref{lemmaB1}}, it follows that  for any $0<\epsilon<1$ there exists a constant $\gamma>0$ such that
\beq
\label{eq_L}
	L_{\epsilon}[\mathcal{Z}^{'}(\Delta)] \leq \gamma \left( \frac{1}{\epsilon} \right)^{k'}.
\eeq
By definition of $\epsilon$-covering, for any ${\bf{z}} \in \mathcal{Z}^{'}(\Delta)$, there exists a vector ${\bf{l}} \in \mathcal{L}_{\epsilon}[\mathcal{Z}^{'}(\Delta)]$ such that 
\beq
\| {\bf{z}}-{\bf{l}} \| \leq \epsilon.
\eeq
Letting ${\bf{v}} = {\bf{z}}-{\bf{l}}$, we have
\begin{align}
	\|{\rm{P}}{\bf{z}}\| &= \| {\rm{P}}({\bf{l}} + {\bf{v}} )\| \nonumber \\
	& \geq \|{\rm{P}}{\bf{l}}\| - \| {\rm{P}}{\bf{v}}\|  \nonumber \\
	& \geq \|{\rm{P}}{\bf{l}}\| - \epsilon,
	\label{eq:ne}
\end{align}
where the last inequality follows from
\begin{align}
 \| {\rm{P}}{\bf{v}} \| &\leq \|{\rm{P}}\| \|\bf{v}\| \nonumber \\
 &= \|{\bf{v}}\|  \nonumber \\
 & \leq \epsilon. 
 \end{align}
 From \eqref{eq:ne}  we have that
if  for all ${\bf{l}} \in \mathcal{L}_{\epsilon}[\mathcal{Z}^{'}(\Delta)]$  we have $\|{\rm{P}}{\bf{l}}\| > 2\epsilon$, then we also have $\|{\rm{P}}{\bf{z}}\|> \epsilon = \epsilon \|{\bf{z}}\|$, and letting $c = \epsilon$ the result follows. What remains to be shown then, is that  for almost every projection ${\rm{P}}$ of rank $m$,  and for all ${\bf{l}} \in \mathcal{L}_{\epsilon}[\mathcal{Z}^{'}(\Delta)]$, we have $\|{\rm{P}}{\bf{l}}\| > 2\epsilon$.

We let
\beq
 	\mathcal{L}_{\epsilon}[\mathcal{Z}^{'}(\Delta)] = \{ {\bf{l}}_1, \cdots, {\bf{l}}_L \},
 \eeq
where $L=L_{\epsilon}[\mathcal{Z}^{'}(\Delta)]$, and  for all $1 \leq i \leq L$ we define 
\beq
	\mathcal{H}_i = \{ {\rm{P}} \in \mathcal{G} : \|{\rm{P}} {\bf{l}}_i \| \leq 2\epsilon \}.
\eeq
 We also let
\beq
	\mathcal{H} = \bigcup_{i=1}^{L} \mathcal{H}_i,
\eeq
so that
\beq
	\mu(\mathcal{H}) = \mu \left(\bigcup_{i=1}^{L} \mathcal{H}_i \right) \leq \sum_{i=1}^{L} \mu(\mathcal{H}_i).
\eeq

We claim that if $\|{\rm{P}}{\bf{l}}\| \leq 2 \epsilon$, then $0 \in {\rm{P}} \mathcal{B}_{2\epsilon}^{\bf{l}}$, where $\mathcal{B}_{2\epsilon}^{\bf{l}} $ is a $2\epsilon$-ball whose center is $\bf{l}$. This can be shown as follows: let ${\bf{b}} = {\bf{l}} - {\rm{P}}{\bf{l}}$, then ${\bf{b}} \in \mathcal{B}_{2\epsilon}^{\bf{l}}$ and ${\rm{P}} {\bf{b}} = {\rm{P}}{\bf{l}} - {\rm{P}}^2 {\bf{l}} = 0$. It follows that
\bea
	\mu({\mathcal{H}_i})
	& \leq & \mu \left( \{ {\rm{P}} \in \mathcal{G} : {0 \in \rm{P}} \mathcal{B}_{2\epsilon}^{{\bf{l}}_i}  \} \right) \nonumber
	\\
	& = & \mu (S(\mathcal{B}_{2\epsilon}^{{\bf{l}}_i})) \nonumber
	\\
	& \leq & \delta \left( 2\epsilon \right)^m, 
\eea
where the last inequality follows from Lemma {\ref{lemmaB2}}. We now have
\bea
	\mu({\mathcal{H}}) 
	&\leq& \sum_{i=1}^{L} \mu(\mathcal{H}_i) \nonumber
	\\
	&\leq& L \delta (2\epsilon)^{m}	\nonumber
	\\
	&\leq& \gamma \delta 2^m \epsilon^{m-k'},
\eea
where the last inequality follows from (\ref{eq_L}). By taking a sufficiently small $\epsilon$, we can now make $\mu({\mathcal{H}})$ arbitrary close to 0, and the proof is complete.

\end{IEEEproof}

By Lemma~\ref{lemmaB3}, there exists a projection $\rm{P}$ of rank $m$ such that   for all ${\bf{z}} \in \mathcal{Z}^{'}(\Delta)$ we have $\|{\rm{P}}{\bf{z}}\| > c \|{\bf{z}}\|$.
By applying Gaussian elimination to such a projection and selecting the non-zero rows of it, we  obtain an $m \times N$ matrix $\rm{A}$. Since $\|{\rm{P}}{\bf{z}}\| = \|{\rm{A}}{\bf{z}}\|$, it follows that  any ${\bf{x}}\in \overline{\mathcal{X}}(\Delta)$ can be robustly recovered from ${\bf{y}} = \rm{A}{\bf{x}} + {\bf{e}}$ with a number of measurements larger than $k'$. 

What remains to be done  is to show  that $k' = \mbox{\normalfont{dim}}_F[\mathcal{Z}^{'}(\Delta)] \leq \mbox{ \normalfont{dim}}_F[\mathcal{X}(\Delta) \oplus \mathcal{X}(\Delta)]$. Let $\mathcal{Z}(\Delta) \subset \overline{\mathcal{Z}}(\Delta)$ containing all elements of $\overline{\mathcal{Z}}(\Delta)$ that have norm at most one.
Since $\mathcal{Z}^{'}(\Delta) \subset \mathcal{Z}(\Delta)$, we  have $k' \leq \mbox{\normalfont{dim}}_F[\mathcal{Z}(\Delta)]$. It is then enough to show that $\mbox{\normalfont{dim}}_F[\mathcal{Z}(\Delta)] = \mbox{ \normalfont{dim}}_F[\mathcal{X}(\Delta) \oplus \mathcal{X}(\Delta)]$.

\begin{lemma}
\label{lemmaB4}
We have
\beq
\label{B4}
	\mbox{\normalfont{dim}}_F[\mathcal{Z}(\Delta)] = \mbox{ \normalfont{dim}}_F[\mathcal{X}(\Delta) \oplus \mathcal{X}(\Delta)]
\eeq
\end{lemma}
\begin{IEEEproof}
Let  ${\bf{z}} \in \mathcal{Z}(\Delta)$ be a vector of  coefficients  of a multi-band function $f_{\bf{z}}$ whose spectral support is bounded by $4 \Omega'$ and whose energy is bounded by one. It follows that $f_{\bf{z}}$ can be represented as 
\beq
	f_{\bf{z}} = f_{{\bf x}_1} + f_{ {\bf x}_2}
\eeq 
where $f_{ {\bf x}_i}, i \in \{ 1,2  \}$ is a multi-band signal whose   spectral support is bounded by $2 \Omega'$ and whose energy is bounded by one. Let ${\bf x}_i$ be a vector of coefficients for $f_{{\bf x}_i}$, $i\in \{1,2\}$. Then, we have 
\beq
	{\bf z} = {\bf x}_1 + {\bf x}_2
\eeq
where ${\bf x}_i \in \mathcal{X}(\Delta)$. Since $\mathcal{Z}(\Delta) \subset \mathcal{X}(\Delta) \oplus \mathcal{X}(\Delta)$, we   conclude that
\beq
	\mbox{\normalfont{dim}}_F[\mathcal{Z}(\Delta)] \leq \mbox{ \normalfont{dim}}_F[\mathcal{X} (\Delta) \oplus \mathcal{X}(\Delta)].
\label{eq_a}
\eeq

Conversely, let us consider ${\bf x}_1, {\bf x}_2 \in \mathcal{X}(\Delta)$. Then, we have
\beq
	\frac{ {\bf x}_1 +  {\bf x}_2}{2} \in \mathcal{Z}(\Delta),
\eeq
which implies $\mathcal{X}(\Delta) \oplus \mathcal{X}(\Delta) \subset 2 \mathcal{Z}(\Delta)$, where $2 \mathcal{Z}(\Delta)$ indicates the set $\{ 2 {\bf{z}} : {\bf{z}}  \in \mathcal{Z}(\Delta) \}$. Therefore, we  conclude that
\beq
	\mbox{\normalfont{dim}}_F[\mathcal{Z}(\Delta)] \geq \mbox{ \normalfont{dim}}_F[\mathcal{X}(\Delta) \oplus \mathcal{X}(\Delta)].
\label{eq_b}
\eeq

By combining (\ref{eq_a}) and (\ref{eq_b}), we obtain the desired result. 
\end{IEEEproof}

\subsection{Proof of Lemma \ref{lt4}}

If  all vectors  ${\bf{x}} \in \mathcal{X}(\Delta)$ can be recovered from ${\bf{y}}={\rm{A}}{\bf{x}}$ , then all vectors  ${\bf{x}} \in \overline{\mathcal{X}}(\Delta)$ can also be  recovered from ${\bf{y}}={\rm{A}}{\bf{x}}$, and vice versa. In the following, we consider $\mathcal{X}(\Delta)$ rather than $\overline{\mathcal{X}}(\Delta)$.

In order to prove Lemma~\ref{lt4}, 
it is enough to show that
a number of measurements 
\beq
\label{condition}
	m <2 \mbox{ \normalfont{dim}}_F[\mathcal{X}(\Delta)]
\eeq
is not sufficient to recover all the elements of ${\mathcal{X}}(\Delta)$ as $T \rightarrow \infty$.

Let us define the set $\mathcal{W}(\Delta) = \mathcal{X}(\Delta) \oplus \mathcal{X}(\Delta)$.
If all ${\bf{x}} \in \mathcal{X}(\Delta)$ can be recovered from {\bf y}, then ${\rm{A}}$ is a one-to-one map on ${\mathcal{X}}(\Delta)$, and vice versa. 
Also, if ${\rm{A}}$ is a one-to-one map on ${\mathcal{X}}(\Delta)$, then
\beq
\label{ker}
	\mbox{ker} ({\rm{A}}) \cap \mathcal{W}(\Delta) = \{ {\bf{0}} \},
\eeq
and vice versa, where $\mbox{ker} ({\rm{A}})$ indicates the kernel of ${\rm{A}}$. 
We then need  to show that (\ref{condition}) violates (\ref{ker}). 
For   convenience of notation, we define $k = 2 \mbox{ \normalfont{dim}}_F[\mathcal{X}(\Delta)]$.

Let us assume that $\mathcal{W}(\Delta)$ contains a $k$-dimensional Euclidean ball. 
Note that (\ref{condition})  implies $\mbox{rank}({\rm{A}}) < k$. Since $\mbox{rank}({\rm{A}}) + \mbox{nullity}({\rm{A}}) = N$, it follows that  $\mbox{nullity}({\rm{A}}) > N - k$. 
This means that the dimension of $\mbox{ker}({\rm{A}})$ is larger than $N-k$, which violates (\ref{ker}) because $\mathcal{W}(\Delta)$ contains a $k$-dimensional Euclidean ball. 

It follows that in order to prove Lemma~\ref{lt4}, it is enough to show that $\mathcal{W}(\Delta)$ contains a $k$-dimensional Euclidean ball.
We will show that this is the case when $T \rightarrow \infty$.

We need some additional definitions, followed by a  preliminary result.

\begin{definition} \label{diameter}
\emph{(Diameter)}. 
	For any $\mathcal{S} \subset  \mathbb{R}^N$, we let  
	\beq
		{\rm{diam}} (\mathcal{S}) = \sup_{\bf{x},\bf{y} \in \mathcal{S}} \| {\bf{x}-\bf{y}} \|.
	\eeq
\end{definition}

\begin{definition} \label{hausdorffmeasure}
\emph{(Hausdorff measure)}. 
	Let $\mathcal{U} \subset \mathbb{R}^N$ and $\{ \mathcal{S}_i \}$ be a cover of $\mathcal{U}$ formed by balls of radius $r < \mu$. We let
	\beq
		\zeta_{\mu}^{s} (\mathcal{U}) = \inf_{\{ \mathcal{S}_i \}} \sum_{i} [{\rm{diam}} (\mathcal{S}_i)]^s.
	\eeq
	The $s$-dimensional Hausdorff measure of $\mathcal{U}$ is given by the limit
	\beq
		\zeta^{s} (\mathcal{U}) = \lim_{\mu \rightarrow 0} \zeta_{\mu}^{s} (\mathcal{U}).
	\eeq
	\end{definition}

\begin{definition} \label{hausdorffdimension}
\emph{(Hausdorff dimension)}.
	For any   $\mathcal{U} \subset \mathbb{R}^N$, the Hausdorff dimension of $\mathcal{U}$ is
	\beq
		\dim_{H} (\mathcal{U}) = \sup \{ s \geq 0 : \zeta^{s} (\mathcal{U}) = \infty \}.
	\eeq
\end{definition}

The Hausdorff dimension has the following two important properties, see~\cite{Falconer}.

{\emph{Property 1.}}  \emph{(Unit ball)}. For any integer $d$  such that $0 \leq d \leq N$, the Hausdorff dimension of the unit ball $B^d (0,1) \subset \mathbb{R}^d \subset \mathbb{R}^N$ is $d$.

{\emph{Property 2.}} \emph{(Countable stability)}. Let $\mathcal{U}_i \subset \mathbb{R}^N$. Then, $\dim_H (\bigcup_{i=1}^{\infty} \mathcal{U}_i) = \sup_{i} \{ \dim_H (\mathcal{U}_i) \}$

From these definitions it follows that
\beq
	\dim_H (\mathcal{U}) \leq \dim_F (\mathcal{U}).
\eeq
However, by Lemma~\ref{falconer} below, if a set satisfies a quasi self-similar property, then the Hausdorff dimension is equal to the fractal dimension. 

\begin{definition}\label{selfsimilar}
\emph{(Quasi self-similarity)}
	Let $\mathcal{U} \subset \mathbb{R}^N$. If for all ${\bf{x,y}} \in \mathcal{U} \cap \mathcal{B}$, there exist $a, r_0 >0$ such that for any ball $\mathcal{B}$ with radius $r < r_0$, there is a mapping $\phi: \mathcal{U} \cap \mathcal{B} \rightarrow \mathcal{U}$ satisfying
	\beq
		a \cdot \| {\bf{x}-\bf{y}} \| \leq r \cdot \| \phi({\bf{x}})-\phi({\bf{y}}) \|,
	\eeq
	then, we say that $\mathcal{U}$ is quasi self-similar. 
\end{definition}

\begin{lemma} \emph{\cite[Theorem 3.]{Falconer1989}}
\label{falconer} 
	Let $\mathcal{U}$ be a nonempty compact subset of $\mathbb{R}^N$ that is  quasi self-similar. Then, 
	\beq
	\dim_H (\mathcal{U}) = \dim_F (\mathcal{U}).
	\eeq
\end{lemma}

We  are now ready to show our final step.

\begin{lemma}
\label{lemma_ball}
	For sufficiently large $T$, the set $\mathcal{W}(\Delta)= \mathcal{X}(\Delta) \oplus \mathcal{X}(\Delta)$ contains a $k$-dimensional Euclidean ball.
\end{lemma}
\begin{IEEEproof}
We have
\beq
\label{union}
	\mathcal{X}(\Delta) = \bigcup_{i} \mathcal{X}_i
\eeq
where $\mathcal{X}_i$ is the set of coefficient vectors of all multi-band signals of a fixed sub-band allocation of measure at most $2\Omega'$ and norm at most one. 
Since $\mathcal{X}(\Delta)$ is a countable union,  by Property 2 of the Hausdorff dimension we have
\beq
\label{Heq1}
	\dim_H [\mathcal{X}(\Delta)] = \sup_i \{ \dim_H (\mathcal{X}_i) \}.
\eeq
Since the Hausdorff dimension of $\mathcal{X}_i$ does not depend on $i$,  we also have that for all $i$

\beq
\label{Heq2}
	\dim_H [\mathcal{X}(\Delta)] =  \dim_H (\mathcal{X}_i).
\eeq
Since $\mathcal{X}(\Delta)$ is a nonempty compact subset of $\mathbb{R}^N$ that is also quasi self-similar   with $a=r_0 = 1$ and $\phi({\bf{x}}) = {\bf{x}}/r$, it follows that
\beq
\label{Heq3}
	 \dim_H [\mathcal{X}(\Delta)]= \dim_F [\mathcal{X}(\Delta)].
\eeq
%

Next, we consider two sets of coefficient vectors $\mathcal{X}_1$ and $\mathcal{X}_2$,  whose sub-bands do not have any intersection. We have
\beq
	{\mathcal{X}}_i = \{ {\bf{x}} : {\bf{x}} = \Phi_i {\bf{\alpha}} {\rm{~where~}} 
	{\| {\bf{x}} \|} \leq 1 {\rm{~and~}}
	{\bf{\alpha}} \in \mathbb{R}^{S} \},
\eeq
for $i= 1,2$. 
By the same argument used in the proof of Lemma~\ref{lemmaA4}, it follows that for $T$ large enough 
the columns of $\Phi_1$ and $\Phi_2$ are independent. Also, note that $\mathcal{X}_i$ is an Euclidean ball 
and by Property one of  the Hausdorff dimension  it follows that
$\mathcal{X}_i$ is a $\dim_H (\mathcal{X}_i)$-dimensional Euclidean ball.
Now,
by definition, $\mathcal{W}(\Delta)$ includes $\mathcal{X}_1 \oplus \mathcal{X}_2$, and since $\mathcal{X}_1$ and $\mathcal{X}_2$ are $\dim_H (\mathcal{X}_i)$-dimensional Euclidean balls and the columns of $\Phi_1$ and $\Phi_2$ are independent, it follows that 
$\mathcal{W}(\Delta)$  contains a $2 \dim_H(\mathcal{X}_i)$-dimensional Euclidean ball. 
Using (\ref{Heq2}) and (\ref{Heq3}), it follows that for $T$ large enough $\mathcal{W}_{\Delta}$   contains  a  $2 \dim_F[\mathcal{X}(\Delta)]$ Euclidean ball, or equivalently, a $k$-dimensional Euclidean ball.

    \end{IEEEproof}

\section{Conclusion}
We have investigated the phase-transition threshold of the minimum
measurement rate sufficient for completely blind reconstruction of any multi-band   signal of given spectral support measure. This threshold has been shown to coincide with twice the fractal dimension per unit ambient dimension of the space spanned by the optimal approximation for bandlimited signals. This result provides an operational characterization of the fractal dimension, and parallels an analogous coding theorem     for the compression of discrete-time, analog, i.i.d.\ sources, where the critical threshold is shown to be equal to the information dimension per unit ambient dimension   of the source~\cite{Wu2010,Wu2012}. Advantages of  the deterministic approach include being oblivious to a priori assumptions on the source distribution, and  providing  recovery guarantees for all signals, rather than  for a large fraction of them. In both cases, fundamental limits apply to the asymptotic regime of large signal dimension. In the stochastic case,  probabilistic concentration is achieved exploiting the ergodicity of the process, while in the deterministic case   vanishing error energy is achieved exploiting spectral concentration.  Despite both  results can be viewed at the high level as an instance of dimensional reduction due to regularity constraints,  the tools required in the deterministic setting are quite different   from those used in
traditional information theory, and include machinery from  approximation theory,  and geometry of functional spaces.  The systematic study of these techniques is clearly desirable, and  this recommendation dates back to  Kolmogorov~\cite{kolmo}. Exploiting some of our  recent results~\cite{IT2017}, we have shown that the price to pay to obtain deterministic guarantees of reconstruction for all signals is only a factor of two in the measurement rate, compared to probabilistic reconstruction. It is also the case that     the absence of  additional spectral information such as the one assumed in \cite{Mishali2009,Mishali2010, Davenport2012}, does not lead to any penalty in the measurement rate.  

Practical achievability schemes for blind reconstruction of continuous signals that come close to the information-theoretic optimum remain an open problem, while much progress has been made for both  discrete-time and continuous-time settings, under various assumptions on what   information about the   signal   is available a priori~\cite{Mishali2009,Mishali2010, Davenport2012,Wu2010,Wu2012,montanari}. Another interesting open question is  the determination of the critical threshold for linear approximation schemes. In this case, without any knowledge of  the spectral support  it is not possible to set-up   the eigenvalue equation  leading to  the optimal subspace approximation~\cite{Landau1980}, and the challenge is to infer the basis functions  directly from the measurements. Investigation of sampling schemes for blind reconstruction is also   of interest, due to their relevance  for practical applications. Our results provide an information-theoretic baseline for performance assessment in all of these cases. Finally, extensions to signals of multiple variables would be of interest in various settings, for example in the context of remote sensing.   In this case, a desirable outcome would be the computation of the  fractal dimension of signals radiated from a bounded domain, generalizing the notion of number of degrees of freedom for bandlimited signals studied in~\cite{mio}, to signals that are sparse in both the frequency and the wavenumber spectrum.

\appendix


\subsection{Proofs of (\ref{appendix1}) and (\ref{appendix2})}

First  let us consider (\ref{appendix1}). Since $\mathcal{X}_{\rm{B}} \subset \mathcal{X}_{\rm{B}} \oplus \mathcal{X}_{\rm{B}}$, we have
\beq
\label{aeq1}
	\mbox{dim}_F(\mathcal{X}_{\rm{B}}) \leq \mbox{dim}_F(\mathcal{X}_{\rm{B}} \oplus \mathcal{X}_{\rm{B}}).
\eeq
For any ${\bf{x}} \in \mathcal{X}_{\rm{B}} \oplus \mathcal{X}_{\rm{B}}$, we have ${\bf{x}}/2 \in \mathcal{X}_{\rm{B}}$, or equivalently ${\bf{x}} \in 2 \mathcal{X}_{\rm{B}}$,    where $2 \mathcal{X}_{\rm{B}}$ indicates the set $\{ 2 {\bf{x}} : {\bf{x}}  \in \mathcal{X}_{\rm{B}} \}$. This implies $\mathcal{X}_{\rm{B}} \oplus \mathcal{X}_{\rm{B}} \subset 2 \mathcal{X}_{\rm{B}}$, and we have 
\beq
\label{aeq2}
	\mbox{dim}_F(\mathcal{X}_{\rm{B}}) \geq \mbox{dim}_F(\mathcal{X}_{\rm{B}} \oplus \mathcal{X}_{\rm{B}}).
\eeq
Combining (\ref{aeq1}) and (\ref{aeq2}), we obatin (\ref{appendix1}). 

Next, we  consider (\ref{appendix2}). We let $\mathcal{X}'_{\rm{B}}$ be a set of vectors such that for any ${\bf{x}} = (x_1, \cdots, x_N) \in \mathcal{X}_{\rm{B}}$, ${\bf{x'}}\in \mathcal{X'}$ is the vector of its first $N'$ components, namely  ${\bf{x'}} = (x_1, \cdots, x_{N'})$ where $N' = \Omega T/\pi + o(T)$.  From inequality (137) in Theorem~6 of  \cite{IT2017}, we have
\beq
\label{H1}
	H_{\epsilon} (\mathcal{X}_{\rm{B}}) \geq H_{\epsilon} (\mathcal{X'}_{\rm{B}}).
\eeq
By inequality (99) of Theorem~3 in  \cite{IT2017}, we have
\beq
\label{H2}
	H_{\epsilon} (\mathcal{X'}_{\rm{B}}) \geq N' \left[ \log \left( \zeta(N') \frac{1}{\epsilon} \right) \right],
\eeq
where $\zeta(N')$ is independent of $\epsilon$. Combining (\ref{H1}) and (\ref{H2}), we obtain
\beq
	H_{\epsilon} (\mathcal{X}_{\rm{B}}) \geq
	N' \left[ \log \left( \zeta(N') \frac{1}{\epsilon} \right) \right].
	\label{equlower} 
\eeq
Similarly, by inequality (138) of Theorem~6 in \cite{IT2017} we have
\beq
\label{H3}
	H_{\epsilon}(\mathcal{X}_{\rm{B}}) \leq H_{\epsilon-\mu} (\mathcal{X'}_{\rm{B}}),
\eeq
and using inequality (100) of Theorem~3 in \cite{IT2017}, we have
\beq
\label{H4}
	H_{\epsilon-\mu} (\mathcal{X'}_{B}) \leq N' \log \left( \frac{1}{\epsilon-\mu} \right) +\eta(N'), 
\eeq
where $0 < \mu < \epsilon$, and $\eta(N')$ is independent of $\epsilon$. Combining (\ref{H3}) and (\ref{H4}), we obtain
\beq
	H_{\epsilon} (\mathcal{X}_{\rm{B}}) \leq
	N' \log \left( \frac{1}{\epsilon-\mu} \right) +\eta(N'). 
\eeq
Since $\mu$ can be arbitrarily small and the logarithm is a continuous function, it follows that 
\beq
\label{equupper}
	H_{\epsilon} (\mathcal{X}_{\rm{B}}) \leq
	N' \log \left( 	1/\epsilon  \right)  +\eta(N'). 
\eeq
Putting together (\ref{equlower}) and (\ref{equupper}), we finally obtain
\beq
\begin{cases}
H_{\epsilon} (\mathcal{X}_{\rm{B}}) & \geq N'   \log \left[ \zeta(N') 1/\epsilon \right],  \label{equno} \\
	H_{\epsilon} (\mathcal{X}_{\rm{B}}) &\leq
	N'  \log \left( 1/\epsilon \right) + \eta(N').
\end{cases}
\eeq
Dividing both sides of \eqref{equno}   by $-\log \epsilon$ and taking the limit for $\epsilon \rightarrow 0$,  we have
\beq
	\mbox{dim}_F(\mathcal{X}_{\rm{B}}) = N',
\eeq
so that
\beq
\label{atemp1}
	\lim_{T \rightarrow \infty}  \frac{\mbox{dim}_F(\mathcal{X}_{\rm{B}})}{T} = \Omega/\pi.
\eeq

\bibliographystyle{IEEEtran}
\bibliography{bib_control}

\begin{thebibliography}{100}
%
\bibitem{Jerri}
A. Jerri. ``The Shannon sampling theorem ---its various extensions and applications: a tutorial review." \emph{Proceedings of the IEEE}, 65(11), pp. 1565--1596, 1977.



\bibitem{SlepianS1} D. Slepian. ``Some comments on Fourier analysis, uncertainty and modeling." \emph{SIAM Review}, 25(3), pp. 379-393, 1983.
%
\bibitem{unser} M. Unser. ``Sampling ---50 years after Shannon." \emph{Proceedings  of the IEEE}, 88(4), pp. 569-587, 2000.
%
\bibitem{Flammer} C. Flammer. ``Spheroidal wave functions." \emph{Stanford University Press}, 1957. Reprinted 2005,  \emph{Dover}.
%
\bibitem{pinkus}
A.~Pinkus, ``n-Widths in approximation theory,'' {\em Springer-Verlag}, 1985.
%
\bibitem{Landau1980}
H. J. Landau and H. Widom, ``The eigenvalue distribution of time and frequency limiting," {\em Journal of Mathematical Analysis and Applications,} 77, pp. 469--481, 1980.

\bibitem{Landau1967}
H. J. Landau, ``Necessary density conditions for sampling and interpolation of certain entire functions" {\em Acta Math.},  vol. 117, pp. 37-52, Feb. 1967.
%
\bibitem{Feng1996}
P. Feng and Y. Bresler, ``Spectrum-blind minimum-rate sampling and reconstruction of multiband signals," in {\em Proc. IEEE Int. Conf. Acoustics, Speech, Signal Processing (ICASSP)}, 3, pp. 1688--1691, May 1996.
%
\bibitem{Bresler1996}
Y. Bresler and P. Feng, ``Spectrum-blind minimum-rate sampling and reconstruction of 2-d multiband signals," in {\em Proc. IEEE Int. Conf. Image Process}, vol. 1, pp. 701--704, Sep. 1996.
%
\bibitem{Venkataramani1998}
R. Venkataramani and Y. Bresler, ``Further results on spectrum blind sampling of 2D signals," in {\em Proc. IEEE Int. Conf. Image Process}, vol. 2, pp. 752--756, Oct. 1998.
%
\bibitem{Mishali2009}
M. Mishali and Y. Eldar, ``Blind multi-band signal reconstruction: compressed sensing for analog signals," in {\em IEEE Trans. Signal Process}, 57 (3), pp. 993--1009, 2009.
%
\bibitem{Mishali2010}
M. Mishali and Y. Eldar, ``From theory to practice: sub-Nyquist sampling of sparse wide-band analog signals," in {\em IEEE J. Select. Top. Signal Process.}, 4 (2), pp. 375--391, 2010.
%
\bibitem{Izu2009}
S. Izu and J. Lakey, ``Time-frequency localization and sampling of multiband signals," in {\em Acta Appl. Math.}, 107 (1), pp. 399--435, 2009.
%
\bibitem{Sejdic2008}
E. Sejdic, M. Luccini, S. Primak, K. Baddour, T. Willink, ``Channel estimation using DPSS based frames," in {\em Proc. IEEE Conf. Accoust., Speech, and Signal Processing (ICASSP)},  Las Vegas, Nevada, March 2008.
%
\bibitem{Davenport2012}
M. A. Davenport and  M. B. Wakin, ``Compressive sensing of analog signals using Discrete Prolate Spheroidal Sequences," in {\em Appl. Comput. Harmon. Anal.,}, 33, pp. 438-472, 2012.
%
\bibitem{Wu2010}
Y. Wu and S. Verdu, ``Renyi information dimension: Fundamental limits of almost lossless analog compression.'' {\em IEEE Trans. on Information Theory}, 56(8), pp. 3721-3748, 2010.
%
\bibitem{Wu2012}
Y. Wu and S. Verdu, "Optimal phase transitions in compressed sensing," {\em IEEE Trans. on Information Theory}, 58,(10), pp. 6241-6263, 2012.
%
%
%
%
\bibitem{Falconer}
K. Falconer, ``Fractal geometry: mathematical foundations and applications," {\em John Wiley \& Sons}, 2004.
%
\bibitem{kolmo}
A. N. Kolmogorov and V. M. Tikhomirov, ``$\epsilon$-entropy and $\epsilon$-capacity of
sets in functional spaces,''  in \emph{Uspekhi Matematicheskikh Nauk}, 14(2),
pp. 3-86, 1959. English translation: \emph{American Mathematical Society
Translation Series}, 2(17), pp. 277-364, 1961.
%
%
%
%
%
\bibitem{csbook}
S. Foucart and H. Rauhut,
``A mathematical introduction to compressive sensing,"
\emph{Springer}, New York, 2013.
%
%
\bibitem{Gleichman2011} S. Gleichman and Y. Eldar, ``Blind compressed sensing," in {\em IEEE Trans. on Information Theory,} 57(10), pp. 6958-6975, 2011.
%
\bibitem{Bres} S.  Ravishankar and Y. Bresler,
``Efficient blind compressed sensing using sparsifying transforms
with convergence guarantees and application to MRI," in {\em SIAM J. Imaging Sciences}, 8(4), pp. 2519-2557, 2015.
%
\bibitem{Reny1} A. R\'{e}nyi. On the dimension and entropy of probability distributions. \emph{Acta Mathematica Hungarica}, 10(1Ð2),  pp. 193-215,
%
\bibitem{Landau1985}
H. J. Landau, ``An Overview of Time and Frequency Limiting.'' Fourier techniques and applications,'' J. F. Price (editor), Plenum, New York,   pp. 201-220, 1985.
%
%
\bibitem{Falconer1989}
K.-J. Falconer, ``Dimensions and measures of quasi self-similar sets." {\em Proceedings of the American mathematical society} 106.(2), pp. 543-554, 1989.
%
\bibitem{Olson2002}
E. Olson, ``Bouligand dimension and almost Lipschitz embeddings." {\em Pacific journal of mathematics}, 202.(2), pp. 459-474, 2002
%
\bibitem{Covering}
J.-L. Verger-Gaugry, ``Covering a ball with smaller equal balls in $\mathbb{R}^n$". Discrete and Computational Geometry," Springer Verlag, 33, pp.143-155, 2005.
%
\bibitem{IT2017}
T. J. Lim and M. Franceschetti, ``Information without rolling dice," {\em IEEE Trans. on Information Theory,} 63(3), pp. 1349-1363, 2017.
\bibitem{montanari} D. L. Donoho, A. Javanmard,  A. Montanari. ``Information-Theoretically Optimal Compressed Sensing via Spatial Coupling and Approximate Message Passing".  {\em IEEE Trans. on Information Theory,} 59(11), pp. 7434 - 7464, 2013.

\bibitem{mio} M. Franceschetti. ``On Landau's eigenvalue theorem and information cut-sets,"
\emph{IEEE Trans. Inf. Theory,}  61(9), pp. 5042-5051,  2015.


\end{thebibliography}

\end{document}